\documentclass[aps,pre,twocolumn,floatfix]{revtex4}
\usepackage[utf8]{inputenc}
\usepackage{amsmath}
\usepackage{graphicx}
\usepackage{amssymb}
\usepackage{epsfig}
\usepackage{natbib}
\usepackage{algorithm}

\def\bq{\begin{equation}}
\def\eq{\end{equation}}

\begin{document}

\title{Wavelet-based density estimation for noise reduction\\ in plasma simulations using particles}

\author{Romain Nguyen van yen}
\affiliation{Laboratoire de M\'et\'eorologie Dynamique-CNRS, \'{E}cole Normale Sup\'erieure, Paris, France}
\author{Diego del-Castillo-Negrete}
\affiliation{Oak Ridge National Laboratory, Oak Ridge, Tennessee, USA}
\author{Kai Schneider}
\affiliation{Laboratoire de M\'ecanique, Mod\'elisation et Proc\'ed\'es Propres-CNRS, and Centre de Math\'ematiques et d'Informatique Universit\'e d'Aix-Marseille, France}
\author{Marie Farge}
\affiliation{Laboratoire de M\'et\'eorologie Dynamique-CNRS, \'{E}cole Normale Sup\'erieure, Paris, France }
\author{Guangye Chen}
\affiliation{Oak Ridge National Laboratory, Oak Ridge, Tennessee, USA }

\begin{abstract}
For given computational resources, the accuracy of plasma simulations
using particles is mainly held back by the noise due to limited
statistical sampling in the reconstruction of the particle distribution
function.
A method based on wavelet  analysis is proposed and tested to reduce this noise. The method, known as wavelet based density estimation (WBDE), was previously introduced in the statistical literature to estimate probability densities given a finite number of independent measurements.
Its novel application to plasma simulations can be viewed as a natural extension of the finite size particles (FSP) approach,
with the advantage of estimating more accurately distribution functions that have localized sharp features.
The proposed method preserves the moments of the particle distribution function to a good level of accuracy, 
has no constraints on the dimensionality of the system, 
does not require an a priori selection of a global smoothing scale, and its able to adapt  locally to the smoothness  of the density based on the given discrete particle data.  
Most importantly, the computational cost of the denoising stage is of the same order as one time step of a FSP simulation. The method is compared with a recently proposed proper orthogonal decomposition based method, and it is tested 
with three particle data sets that involve different levels of collisionality and interaction with external and self-consistent fields. 
\end{abstract}

\maketitle

%%%%%%%%%%%%%%%%%%%%%%%%%
\section{Introduction}
%%%%%%%%%%%%%%%%%%%%%%%%%

Particle-based numerical methods are routinely used in plasma physics 
calculations \cite{Birdsall1985,Hockney1988}.  In many cases these methods are 
 more efficient and simpler to implement than the corresponding continuum 
Eulerian methods.  However, particle methods face the well known 
statistical sampling limitation of attempting to simulate a physical system containing $N$ 
particles using  $N_p \ll N$  computational particles. 
Particle methods do not seek to reproduce the exact
individual behavior of the  particles, but rather to approximate 
statistical macroscopic quantities  like density, current, 
and temperature. These quantities are determined from the particle 
distribution function. Therefore, a problem of relevance for 
the success of  particle-based simulations is the reconstruction of the 
particle distribution function from  discrete particle data. 

The difference between the distribution function reconstructed from a 
simulation using $N_p$ particles and the exact distribution function 
gives rise to a discretization error generically known as ``particle noise'' due to its random-like 
character.  Understanding and reducing this error is a complex problem 
of importance in the validation and verification of particle
codes, see for example Refs.~\cite{Nevins2005,Krommes2007,McMillan2008}  and 
references therein for a discussion in the context of 
gyrokinetic calculations. One obvious way to reduce particle noise is by 
increasing the number of computational particles. However, the 
unfavorable scaling of the error with the number of particles,  $\sim 
1/\sqrt{N_p}$ \cite{Krommes1993,Aydemir1994},  puts a severe limitation on this 
straightforward approach.  
This has motivated the development of various noise reduction techniques 
including finite size particles (FSP) \cite{Hockney1966,ABL70}, Monte-Carlo methods \cite{Aydemir1994}, Fourier-filtering 
\cite{Jolliet2007},  coarse-graining \cite{Chen2007}, Krook operators \cite{McMillan2008}, 
smooth interpolation  \cite{Shadwick2008}, low noise 
collision operators \cite{Lewandowski2005}, and Proper Orthogonal 
Decomposition (POD) methods \cite{delCastillo2008} among others.  

In the  present paper we propose a wavelet-based method for noise 
reduction in  the  reconstruction of particle distribution 
functions from  particle simulation data.  
The method, known as  Wavelet Based Density Estimation (WBDE),  was 
originally introduced in Ref.~\cite{Donoho1996}  in the context of statistics
to estimate  probability densities given a finite number of independent 
measurements.  However, to our knowledge, this method has not been 
applied before to particle-base computations. WBDE, as used here, is based on truncations of the wavelet representation of the Dirac 
delta function  associated with each particle. The method  
yields almost optimal results for functions with unknown local smoothness without compromising computational efficiency, assuming that the particles' coordinates are statistically independent.
As a first step in the application of the WBDE method to plasma 
particle simulations,  we limit attention to ``passive denoising''. That 
is  the WBDE method is treated  as a post-processing technique 
applied to independently generated  particle data. The problem of ``active denoising'', e.g. the application 
of WBDE methods in the  evaluation of self-consistent fields in 
particle in cell simulations, will not be addressed.  This 
simplification will allow us to assess the efficiency of the proposed noise 
reduction method in a simple setting.  Another simplification 
pertains the dimensionality. 
Here, for the sake of simplicity,  we limit attention to the reconstruction and denoising problem 
in two dimensions. However, the extension of the WBDE method to 
higher dimensions is in principle straightforward. 

Collisions, or the absence of them, play an important role in plasma 
transport problems.  Particle methods handle the collisional and 
non-collisional parts of the dynamics differently. Fokker-Planck-type collision operators 
are typically introduced in particle methods using Langevin-type 
stochastic differential 
equations. On the other hand, the non-collisional part of the dynamics is described using 
deterministic ordinary differential equations. Collisional dominated 
problems tend to washout small scale structures whereas 
collisionless problems typically  develop fine scale filamentary 
structures in phase space. Therefore, it is important to test the dependence 
of the efficiency of denoising reconstruction methods on the level of 
collisionality. 
%% 
 % A problem of interest is the relationship between the 
 % stochasticity introduced to model collisions and the particle noise  
 % due to statistical sampling errors ($N_p \ll N$). 
 %%
Here we test the WBDE method in strongly collisional, weakly collisional and 
collisionless regimes. For the strongly collisional regime we consider 
particle data of force-free collisional relaxation 
involving energy and pinch-angle scattering.  The weakly collisional 
regime is illustrated using guiding-center particle data of a 
magnetically confined plasma in toroidal geometry. 
The collisionless regime is  studied using particle in cell (PIC) 
data  corresponding to  bump-on-tail and two streams 
instabilities in the Vlasov-Poisson system. 

Beyond the role of collisions,  the data sets that we are considering open the possibility of exploring the role of external and self-consistent fields in the reconstruction of the particle density. 
 In the collisional relaxation problem no forces act on the particles, in the guiding-center problem particles interact with an external magnetic field, and in the Vlasov-Poisson  problem particle interactions are incorporated
through a self-consistent electrostatic mean field.  One of the goals of this paper is to compare the WBDE method with the Proper Orthogonal Decomposition (POD) density reconstruction method proposed in Ref.~\cite{delCastillo2008}.

The rest of the paper is organized as follows. 
In Sect.~II we review the main properties of kernel density estimation (KDE) and show its relationship with finite size particles (FSP). We then review basic notions on orthogonal wavelet and multiresolution analysis and outline a step by step algorithm for WBDE. Also, for completeness, in this section we include a brief description of the POD 
reconstruction method proposed in Ref.~\cite{delCastillo2008}. 
Section~III discusses applications of the WBDE method and the 
comparison with the POD method. 
We start by post-processing a simulation of plasma relaxation by random collisions against a background thermostat.
We then turn to a $\delta f$ Monte-Carlo simulation in toroidal geometry, whose phase space has been reduced to two dimensions.
Finally, we analyze the results of particle-in-cell (PIC) simulations of a 1D Vlasov-Poisson plasma.
The conclusions are presented in Sec.~IV. 

%the Vlasov equation, namely, to follow the time evolution of the particle distribution function $f$, according to:
%\begin{equation}\label{Vlasov}
%\frac{\partial{f}}{\partial t} + \bf{v} \cdot \bf {\nabla}_x f + \frac{\bf{F}}{m} \bf{\nabla}_v f = 0 
%\end{equation}

%%%%%%%%%%%%%%%%%%%%%%%%%
\section{Methods}
%%%%%%%%%%%%%%%%%%%%%%%%%

This section presents the wavelet-based density estimation (WBDE) 
algorithm. We start by reviewing basic ideas on kernel density 
estimation (KDE) which is closely related to the use of finite size 
particles (FSP) in PIC simulations. Following this, we 
we give a brief introduction to wavelet analysis and discuss the WBDE algorithm. 
For completeness, we also include a brief summary of the POD 
approach. 
%%%%%%%%%%%%%%%%%%%
\subsection{Kernel density estimation}
%%%%%%%%%%%%%%%%%%

Given a sequence of  independent and identically distributed  measurements, the nonparametric density estimation problem consists in finding the  underlying probability density function (PDF), with no 
a priori assumptions on its functional form. Here we discuss general ideas  on 
this difficult problem for which a variety of statistical methods have been developed. 
Further details can be found in the statistics literature, e.g. Ref.~\cite{Silverman1986}.

Consider a number $N_p$ of statistically independent particles with phase space coordinates $(\mathbf{X}_n)_{1\leq n \leq N_p}$ distributed in $\mathbb{R}^d$  according to a PDF $f$. This data 
can come  from a PIC or a Monte-Carlo,  full $f$ or $\delta f$ simulation. 
Formally, the sample PDF can be written as
\begin{equation}\label{dirac_estimate}
f^\delta(\mathbf{x}) = \frac{1}{N_p}\sum_{n=1}^{N_p} \delta(\mathbf{x}-\mathbf{X}_n)
\end{equation}
where $\delta$ is the Dirac distribution.
Because of  its lack of smoothness,  Eq.~(\ref{dirac_estimate}) is far from the actual distribution $f$ according to most reasonable definitions of the error. Moreover, the dependence of  $f^\delta$  on the statistical fluctuations in $(\mathbf{X}_n)$ can lead to an artificial increase of the collisionality of the plasma.

The simplest method to introduce some smoothness in $f^\delta$ is to use a histogram.
Consider a tiling of the phase space by a Cartesian grid with $N_g^d$ cells. Let  $\left\{B_\lambda\right\}_{\lambda\in\Lambda}$ denote 
the set of all cells with characteristic function
 $\chi_\lambda$  defined as  $\chi_\lambda=1$ if $x \in B_\lambda$ and  $\chi_\lambda=0$ otherwise. 
Then the histogram corresponding to the tiling is
\begin{equation}\label{histogram_estimate}
f^H(\mathbf{x}) = \sum_{\lambda \in \Lambda} \left(\frac{1}{N_p}\sum_{n=1}^{N_p} \chi_\lambda(\mathbf{X}_n) \right) \chi_\lambda(\mathbf{x})
\end{equation}
which can also be viewed as the orthogonal projection of $f^\delta$ on the space spanned by the $\chi_\lambda$. The main difference between $f^{\delta}$ and $f^H$ is that the latter cannot vary at scales finer than the grid scale which is of order $N_g^{-1}$.
By choosing $N_g$ small enough, it is therefore possible to reduce the variance of $f^H$ to very low levels, 
but the estimate then becomes more and more biased towards a piecewise continuous function, which is not smooth enough to be the true density. Histograms correspond to the nearest grid point (NGP) charge assignment scheme  used in the early  days of plasma physics computations \cite{Hockney1966}.

One of the most  popular methods
to achieve higher level of smoothness is kernel density estimation (KDE) \cite{Parzen1962}.
Given  $(\mathbf{X}_n)_{1\leq n \leq N_p}$, the kernel estimate of 
$f$ is defined as
\begin{equation}\label{kernel_estimate}
f^K(\mathbf{x}) = \frac{1}{N_p}\sum_{n=1}^{N_p} K(\mathbf{x}-\mathbf{X}_n)\, ,
\end{equation}
where the smoothing kernel $K$ is a positive definite, normalized, $\int K =1$, function.
Equation~(\ref{kernel_estimate}) corresponds to the convolution of $K$ with the Dirac delta measure corresponding to each particle. 
A typical example is the Gaussian kernel 
\begin{equation}\label{gaussian_kernel}
K_h(\mathbf{x}) = \frac{1}{(\sqrt{2\pi}h)^d} e^{-\frac{\Vert \mathbf{x} \Vert^2}{2h^2}}
\end{equation}
where the  so-called ``bandwidth'', or smoothing scale, $h$, is a free parameter. 
The optimal smoothing scale depends on how the error is measured. 
For example, in the one dimensional case, to minimize the mean $L^2$-error between the estimate and the true density, the smoothing volume $h^d$ should scale like ${N_p}^{-\frac{1}{5}}$, and the resulting error scales like $N_p^{-\frac{2}{5}}$ \cite{Silverman1986}.
As in the case of  histograms, the choice of $h$ relies on a trade-off between variance and bias.
In the context of plasma physics simulations the kernel $K$ corresponds to the charge assignment function  \cite{Hockney1988}.

A significant effort has been devoted in the choice of the function $K$ since it has a strong impact on computational efficiency and on the conservation of global quantities. Concerning $h$, it has been shown that it should not be much larger than the Debye length $\lambda_D$ of the plasma to obtain a realistic and stable simulation \cite{Birdsall1985}.
Given a certain amount of computational resources, the general tendency has thus been to reduce $h$ as far as possible in order to fit more Debye lengths inside the simulation domain,
which means that the effort has been concentrated on reducing the bias term in the error.
Since the force fields depend on $f$ through integral equations,
like the Poisson equation, that tend to reduce the high wavenumber noise,
we do not expect the disastrous scaling $h \propto {N_p}^{-\frac{1}{5}}$,
which would mean $N_p \propto \lambda_D^{5d}$ in $d$ dimensions, to hold.
Nevertheless, the problem remains that if we want to preserve high resolution features of $f$ or of the electromagnetic fields, we need
to reduce $h$, and therefore greatly increase the number of particles to prevent the simulation from drowning into noise.
Bandwidth selection has long been recognized as the central issue in kernel density estimation \cite{Chiu1991}.
We are not aware of a theoretical or numerical prediction of the optimal value of $h$ taking into account the noise term. 
To bypass this difficulty, it is possible to use new statistical methods which do not force us to choose a global smoothing parameter.
Instead, they adapt locally to the behavior of the density $f$ based on the available data.
Wavelet based-density estimation, which we will introduce in the next two sections, is one of these methods.

% The same kind of arguments, interpreted from a physical point of view, have led to the introduction of finite size particles (FSP) in plasma simulations \cite{ABL70}.
% This is not surprising since both methods have the same goal, namely to accelerate the convergence of the empirical PDF to the PDF of the underlying model.
% The FSP equivalent of the bandwidth is the so-called particle size.
% The situation is more complicated in the case of $\delta f$ codes, since only a small part of the distribution function is then approximated using particles : 

%%%%%%%%%%%%%%%%%
\subsection{Bases of orthogonal wavelets}
%%%%%%%%%%%%%%%%%

Wavelets are a standard mathematical tool to analyze and compute non stationary signals. Here  we recall basic concepts and definitions. Further details can be found in Ref.~\cite{MF92} and references therein. The construction takes place in the Hilbert space 
$L^2(\mathbb{R})$ 
of square integrable functions.
%on $\mathbb{R}$.
An orthonormal family $(\psi_{j,i}(x))_{j\in\mathbb{N},i\in\mathbb{Z}}$ is called a wavelet family when its members are dilations and translations of a fixed function $\psi$ called the mother wavelet:
\begin{equation}\label{wavelet_1}
 \psi_{j,i}(x) = 2^{j/2}\psi(2^j x-i)
\end{equation}
where $j$ indexes the scale of the wavelets and $i$ their positions, and  $\psi$ satisfies $\int \psi = 0$.
In the following we shall always assume that $\psi$ has compact support of length $S$.
The coefficients $\langle f \mid \psi_{j,i} \rangle = \int f \psi_{j,i} $ of a function $f$ for this family are denoted by $(\tilde{f}_{j,i})$.
These coefficients describe the fluctuations of $f$ at scale $2^{-j}$ around position $\frac{i}{2^j}$.
Large values of $j$ correspond to fine scales, and small values to coarse scales.
Some members of the commonly used Daubechies 6 wavelet family
are shown in the left panel of  Fig.~1. 
%Wavelets at scale $j=1$ will have compact support of length $\frac{S}{2}$, $\frac{S}{4}$ for $j=2$, and so on.
%The $2^{j/2}$ prefactor ensures proper $L^2$ normalization of the wavelets despite their shrinking support.

It can be shown that the orthogonal complement in $L^2(\mathbb{R})$ of the linear space spanned by the wavelets
is itself orthogonally spanned by the translates of a function $\varphi$, called the scaling function.
%:
%\begin{equation}
% \varphi_{0,i}(x) = \varphi(x-i)
%\end{equation}
% and the coefficients $\langle f \mid \varphi_{i} \rangle$ are denoted $\bar{f}_{0,i}$.
%In fact, one can also shrink the scaling function $\varphi$ :
Defining
\begin{equation}\label{phi_shrink}
\varphi_{L,i} = 2^{\frac{L}{2}} \varphi(2^L x - i)
\end{equation}
and the scaling coefficients $\bar{f}_{L,i} = \langle f \mid \varphi_{L,i} \rangle$, 
one thus has the reconstruction formula:
\begin{equation}\label{wavelet_reconstruction}
f = \sum_{i=-\infty} ^{\infty} \bar{f}_{L,i} \varphi_{L,i} + \sum_{j=L}^{\infty}\sum_{i=-\infty}^{\infty} \tilde{f}_{j,i} \psi_{j,i}
\end{equation}
The first sum on the right hand side of Eq.~(\ref{wavelet_reconstruction}) is a smooth approximation of $f$ at the coarse scale, $2^{-L}$, and the second sum corresponds to the addition of details at successively finer scales. 

If the wavelet $\psi$ has $M$ vanishing moments:
\begin{equation}\label{vanishing_moments}
\int x^m \psi(x) dx = 0
\end{equation}
for $0 \leq m < M$, and if $f$ is locally $m$ times continuously differentiable around some point $x_0$,
then a key property of the wavelet expansion is that the coefficients located near $x_0$ decay when $j\to\infty$ like $2^{-j(m+\frac{1}{2})}$ \cite{Jaffard1991}.
Hence, localized singularities or sharp features in $f$ affect only a finite number of wavelet coefficients within each scale.
Another important consequence of (\ref{vanishing_moments}) of special relevance to particle methods
 is that for $0 \leq m < M$, the moments $\int x^m f(x) dx$ of the particle distribution function depend only on its scaling coefficients, and not on its wavelet coefficients.
 
If the scaling coefficients $\overline{f}_{J,i}$ at a certain scale $J$ are known, 
all the wavelet coefficients at coarser scales ($j \leq J$) can be computed using the fast wavelet transform (FWT) algorithm \cite{SM00}.
We shall address the issue of computing the scaling coefficients themselves in section \ref{critical_discussion}.

\begin{figure}
\includegraphics[width=0.5\columnwidth,height=0.4\columnwidth]{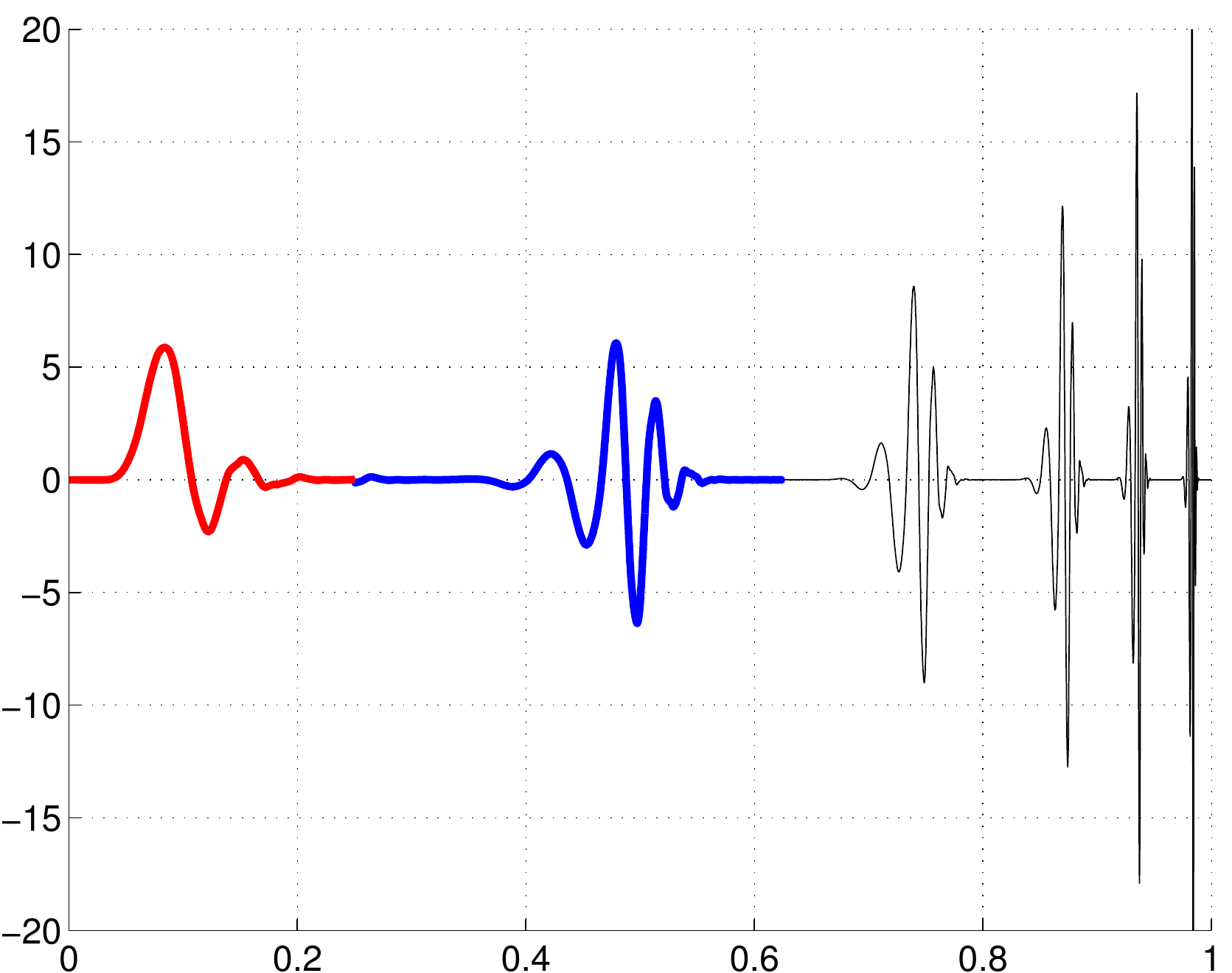}
\hspace{0.05cm}
\includegraphics[width=0.44\columnwidth]{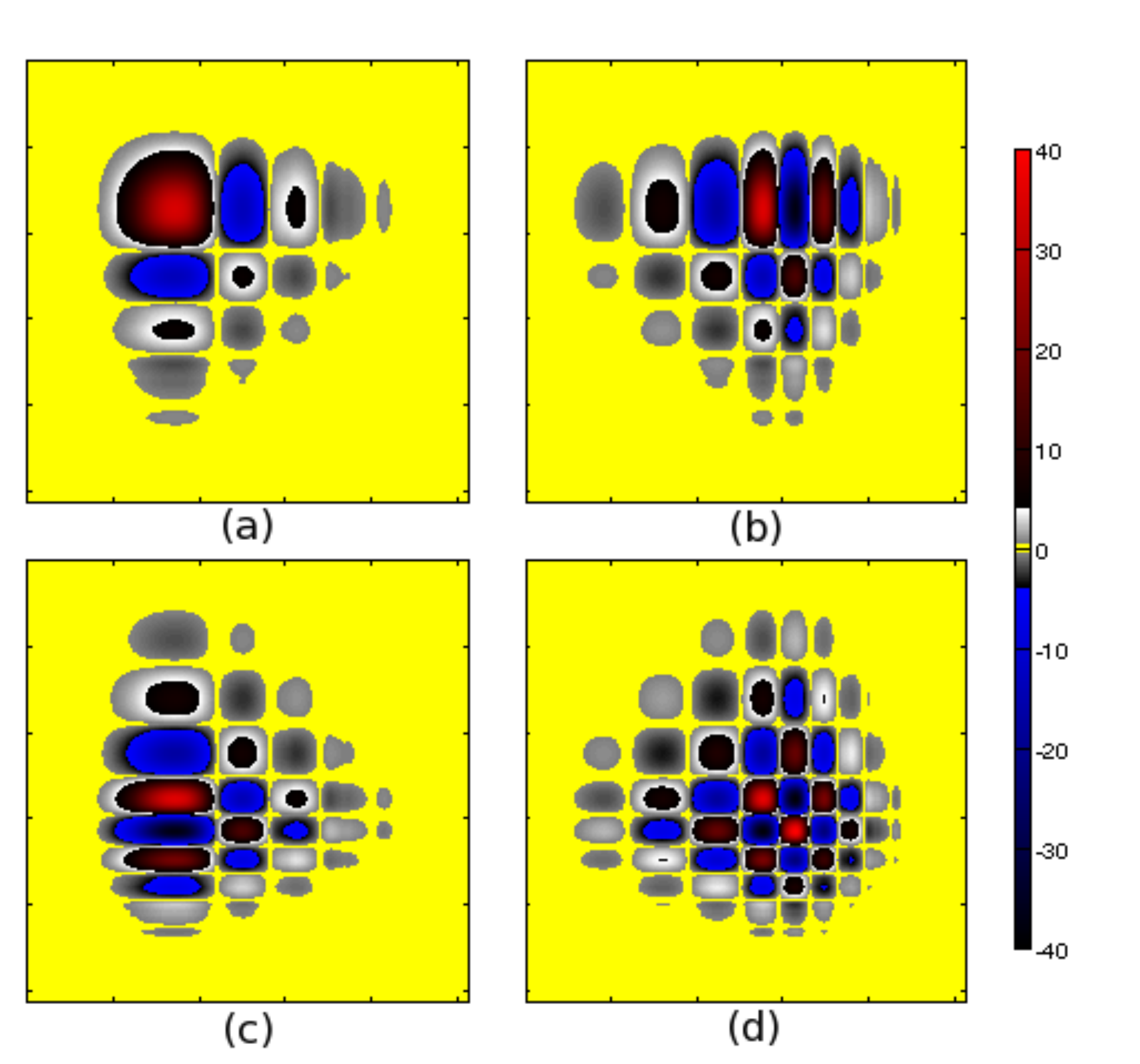}
\caption{
\label{daubechies_wavelets_1D}Daubechies 6 wavelet family.
Left, bold red: scaling function $\varphi$ at scale $j = 5$.
Left, bold blue: wavelet $\psi$ at scale $j = 5$.
Left, thin black, from left to right: wavelets at scales 6, 7, 8 and 9.
Right : (a) 2D scaling function $\varphi(x_1)\varphi(x_2)$. 
(b) first 2D wavelet $\psi(x_1)\varphi(x_2)$.
(c) second 2D wavelet $\varphi(x_1)\psi(x_2)$.
(d) third 2D wavelet $\psi(x_1)\psi(x_2)$.
}
\end{figure}

The generalization to $d$ dimensions involves tensor products of wavelets and scaling functions at the same scale.
For example, given a wavelet basis on $\mathbb{R}$, a wavelet basis on $\mathbb{R}^2$ can be constructed in the following way:
\begin{eqnarray}
\psi^1_{j,i_1,i_2}(x_1,x_2) &=& 2^j\psi(2^j x_1-i_1) \varphi(2^j x_2-i_2) \\
\psi^2_{j,i_1,i_2}(x_1,x_2) &=& 2^j\varphi(2^j x_1-i_1) \psi(2^j x_2-i_2) \\
\psi^3_{j,i_1,i_2}(x_1,x_2) &=& 2^j\psi(2^j x_1-i_1) \psi(2^j x_2-i_2 ) \, ,
\end{eqnarray}
where we refer to the exponent $\mu = 1, 2, 3$ as the direction of the wavelets.
This name is easily understood by looking at different wavelets shown in Fig.~\ref{daubechies_wavelets_1D} (right).
The corresponding scaling functions are simply given by $2^j \varphi(2^j x_1 - i_1)\varphi(2^j x_2 - i_2)$.
Wavelets on $\mathbb{R}^d$ are constructed exactly in the same way, but this time using $2^d-1$ directions.
To lighten the notation we write the $d$-dimensional analog of Eq.~(\ref{wavelet_reconstruction}) as
\begin{eqnarray}\label{wavelet_reconstruction_2D}
 f = \sum_{\lambda\in\Lambda_{\phi,L}} \overline{f}_\lambda \phi_\lambda + \sum_{\lambda\in\Lambda_{\psi,L}} \tilde{f}_\lambda \psi_\lambda
\end{eqnarray}
where $\lambda = (j,\mathbf{i},\mu)$ is a multi-index,
with the integer $j$ denoting the scale and the  integer vector $\mathbf{i} = (i_1,i_2,\ldots)$ denoting the 
position of the wavelet.

The wavelet multiresolution reconstruction formula in Eq.~(\ref{wavelet_reconstruction}) involves an infinite sum over the 
position index $i$. One way of dealing with this sum is to determine a priori the non-zero coefficients in Eq.~(\ref{wavelet_reconstruction}),
and work only with these coefficients, but still retaining the full wavelet basis on $\mathbb{R}^d$ as presented above.
%Indeed, in what we are doing, since there are finitely many particles and the wavelets have compact support,
%there will be only finitely many non-zero wavelet coefficients and scaling coefficients.
Another  alternative, which we have chosen because it is easier to implement, is to periodize the wavelet transform on a bounded domain \cite{SM00}.
Assuming that the coordinates have been rescaled so that all the particles lie in $[0,1]^d$,
we replace the wavelets and scaling functions by their periodized counterparts:
\begin{eqnarray}\label{periodized_wavelets}
\psi_{j,i}(x) & \to & \sum_{l=-\infty}^{\infty} \psi_{j,i}(x+l) \\
\varphi_{j,i}(x) & \to & \sum_{l=-\infty}^{\infty} \varphi_{j,i}(x+l) \, .
\end{eqnarray}
Throughout this paper we will consider only periodic wavelets. 
For the sake of completeness we mention a third alternative which is technically more complicated. 
It consists in constructing a wavelet basis on a bounded interval \cite{Cohen1993}.
The advantage of this approach is that it does not introduce artificially large wavelet coefficients at the boundaries for functions $f$ that 
are not periodic.

%%%%%%%%%%%%%%%%%
\subsection{Wavelet based density estimation}
%%%%%%%%%%%%%%%%%

The multiscale nature of wavelets allows them to adapt locally to the smoothness of the analyzed function \cite{SM00}.
This fundamental property has triggered their use in a variety of problems.
One of their most fruitful applications has been the denoising of intermittent signals \cite{DJ94}.
The practical success of wavelet thresholding to reduce noise relies on the observation that the expansion of  signals in a wavelet basis is typically sparse.
Sparsity means that the interesting features of the signal are well summarized by a small fraction of large wavelet coefficients.
On the contrary, the variance of the noise is spread over all the coefficients appearing in Eq.~(\ref{wavelet_reconstruction_2D}).
Although the few large coefficients are of course also affected by noise, curing the noise in the small coefficients is already a very good improvement.
The original setting of this technique, hereafter referred to as global wavelet shrinkage, requires the noise to be additive, stationary, Gaussian and white.
It found a first application in plasma physics in Ref.~\cite{Farge2006},  where coherent bursts were extracted out of plasma density signals.
Since Ref.~\cite{DJ94}, wavelet denoising has been extended to a number of more general situations, like non-Gaussian or correlated additive noise, or to denoise the spectra of locally stationary time series \cite{Sachs1996}.
In particular, the same ideas were developed in Ref.~\cite{Vannucci1995,Donoho1996} to propose a wavelet-based density estimation (WBDE) method based on independent observations.
At this point we would like to stress that WBDE assumes nothing about the Gaussianity of the noise or whether or not it is stationary.
In fact, under the independence hypothesis -- which is admittedly quite strong -- the statistical properties of the noise are entirely determined by standard probability theory.
We refer to Ref.~\cite{Vidakovic1999} for a review on the applications of wavelets in statistics.
In Ref.~\cite{Gassama2007}, global wavelet shrinkage was applied directly to the charge density of a 2D PIC code, in a case were the statistical fluctuations were quasi Gaussian and stationary.
In particular, an iterative algorithm \cite{AAMF04}, which crucially relies on the stationnarity hypothesis, was used to determine the level of fluctuations. However,in the next section we will show an example where the noise is clearly non-stationary, and  this procedure fails.

Let us now describe the WBDE method as we have generalized it to several dimensions.
The first step is to  expand the sample particle distribution function, $f^\delta$, in Eq.~(\ref{dirac_estimate})  in a wavelet basis
according to Eq.~(\ref{wavelet_reconstruction_2D}) with the wavelet coefficients 
\begin{eqnarray}\label{empirical_wavelet_coefficients}
 \overline{f}_{\lambda} & = & \langle f^\delta \mid \varphi_\lambda \rangle = \frac{1}{N_p} \sum_{n=1}^{N_p} \varphi_\lambda(X_n) \\
 \tilde{f}_{\lambda} & = & \langle f^\delta \mid \psi_\lambda \rangle = \frac{1}{N_p} \sum_{n=1}^{N_p} \psi_\lambda(X_n) \, .
\end{eqnarray}
Since this reconstruction is exact, keeping all the wavelet coefficients does not improve the smoothness of $f^\delta$.
The simple and yet efficient remedy consists in keeping only a subset of the wavelet coefficients in Eq.~(\ref{wavelet_reconstruction_2D}).
A straightforward prescription would be to discard all the wavelet coefficients at scales finer than a cut-off scale $L$.
This approach corresponds to a  generalization of the histogram method in Eq.~(\ref{histogram_estimate}) with $N_g = 2^L$. Because the characteristic functions $\chi_\lambda$ of the cells in a dyadic grid are the scaling functions associated with the Haar wavelet family, Eqs.~(\ref{wavelet_reconstruction_2D}) and (\ref{histogram_estimate}) are in fact equivalent for this wavelet family.  Accordingly, like in the histogram case, we would have to choose $L$ quite low to obtain a stable estimate, at the risk of losing some sharp features of $f$.
Better results can be obtained by keeping some wavelet coefficients down to a much finer scale $J > L$.
However, to prevent that statistical fluctuations contaminate the estimate, only those coefficients whose modulus are above a certain threshold should be kept.
We are thus naturally led to a nonlinear thresholding procedure.
In the one dimensional case, values of $J$, $L$, and of the threshold within each scale that yield theoretically optimal results have been given in Ref.~\cite{Donoho1996}. 
This reference discusses the precise smoothness requirements on $f$,  which can accommodate well localized singularities, like shocks and filamentary structures known to arise in collisionless plasma simulations.
There remains the question of how to compute the $\tilde{f}_{j,i}$ based on the positions of the particles.
Although more accurate methods based on (\ref{empirical_wavelet_coefficients}) may be developed in the future, our present approximation relies on the computation of a histogram, which creates errors of order $N_g^{-1}$.
The complete procedure is described in the following {\bf 
Wavelet-based density estimation} algorithm:
%\hline
%\begin{algorithm}
%\caption{
%\label{WBDE_algo}
%Wavelet-based density estimation
%}
%\rule[-0.1cm]{5cm}{0.01cm}
\begin{enumerate}
\item \label{histogram_approx} construct a histogram $f^H$ of the particle data with $N_g = 2^{J_g}$ cells in each direction,
\item \label{scaling_function_approx} approximate the scaling coefficients at the finest scale $J_g$ by :
\begin{equation}\label{scaling_function_approximation}
 \overline{f}_{J_g,\mathbf{i}} \simeq 2^{-{J_g}/{2}} f^H(2^{-J_g} \mathbf{i})
\end{equation}
 \item compute all the needed wavelet coefficients using the FWT algorithm,
\item keep all the coefficients for scales coarser than $L$, defined by $2^{dL}\sim N_{p}^{\frac{1}{1+2r_{0}}}$ where $r_{0}$ is the order of regularity of the wavelet (1 in our case),
\item discard all the coefficients for scales strictly finer than $J$ defined by $2^{dJ}\sim\frac{N_{p}}{\log_{2}N_{p}}$,
\item \label{thresholding_function} for scales $j$ in between $L$ and $J$, keep only the wavelet coefficients $\tilde{f}_{\lambda}$ such that $\vert\tilde{f}_{\lambda}\vert\geq T_j = C\sqrt{\frac{j}{N_{p}}}$ where $C$ is a constant that must in principle depend on the smoothness of $f$ and on the wavelet family \cite{Donoho1996}.
\end{enumerate}
%\rule[-0.1cm]{5cm}{0.01cm}
%\end{algorithm}

In the following, except otherwise indicated,  $C=\frac{1}{2}$. For the wavelet bases we 
used  orthonormal Daubechies wavelets with 6 vanishing moments and thus support of size $S=12$ \cite{Daubechies1992}.
In our case, $r_0 = 1$, which means that the wavelets have a first derivative but no second derivative, and the size of the wavelets at scale $L$ for $d=1$ is roughly $N_p^{-\frac{1}{3}}$.
%This corresponds to the bandwidth that would have been optimal in the mean square sense for a kernel estimate with the same number of particles.
Since $N_p \gg 1$, it follows  from the definition at stage 5 of the algorithm that the size of the wavelets at scale $J$ is orders of magnitude smaller than that.
Using the adaptive properties of wavelets, we are thus able to detect small scale structures of $f$ without compromising the stability of the estimate.
Note that the error at stage \ref{scaling_function_approx} could be reduced by using Coiflets \cite{Daubechies1993} instead of Daubechies wavelets, 
but the gain would be negligible compared to the error made at stage \ref{histogram_approx}.
We will denote the WBDE estimate of $f$ as $f^W$. 
In the one-dimensional case, 
\begin{equation}\label{explicit_wavelet_estimate}
{f^{W}}=\sum_{i=1}^{2^{L}}\overline{f}_{L,i}\varphi_{L,i}+\sum_{j=L}^{J}\sum_{i=1}^{2^{j}}\tilde{f}_{j,i} \rho_{j}(\tilde{f}_{j,i})\psi_{j,i}
\end{equation}
where $\rho_{j}$ is the thresholding function as defined by stage \ref{thresholding_function} of the algorithm : $\rho_{j}(y) = 0$ if $\vert y \vert \leq T_j$ and $\rho_{j}(y) = 1$ otherwise.

Finally, let us propose two methods for applying WBDE to postprocess $\delta f$ simulations.
Recall that the Lagrangian equations involved in the $\delta f$ schemes are identical to their full $f$ counterparts.
The only difficulty introduced by the $\delta f$ method lies in the evaluation of phase space integrals of the form  $\delta I = \int A \cdot (f-f_0)$,
where $A$ is a function on phase space and $f_0$ is a known reference distribution function.
In these integrals, $f-f_0$ should be replaced by $\delta f$, which is in turn written as a product $wf$, where $w$ is a ``weighting'' function.
%Despite its name, $w$ is usually not a positive function.
Numerically, $w$ is known via its values at particles positions, $w(X_n)$, and the usual expression for $\delta I$ is thus $\delta I = \sum_{n=1}^{N_p} A(X_n) w(X_n)$.
We cannot apply WBDE directly to $\delta f$, since this function is not a density function.% and the thresholds $T_j$ therefore do not apply.
An elegant approach would be to first apply WBDE to the unweighted distribution $f^\delta$ to determine the set of statistically significant wavelet coefficients, and to include the weights only in the final reconstruction (\ref{explicit_wavelet_estimate}) of $f^W$.
%This amounts to replacing the first occurence of $\tilde{f}_{j,i}$ by $\tilde{\delta f}_{j,i}$ in , and leaving the second occurence intact.
A simpler approach, which we will illustrate in section \ref{delta_5d_example}, consists in renormalizing $\delta f$, so that 
$%\begin{equation}\label{delta_f_normalization}
\int\vert\delta f\vert = 1$, 
and treat it like a density.

%%%%%%%%%%%%%%%%%
\subsection{Further issues related to practical implementation}
\label{critical_discussion}
%%%%%%%%%%%%%%%%%

In this section we discuss how the WBDE method handles two issues of direct relevance to plasma simulations:
conservation of moments and computational efficiency.
As mentioned before, due to the vanishing moments of the wavelets in Eq.~(\ref{vanishing_moments}),
the moments up to order $M$ of the particle distribution distribution are solely determined by its scaling function coefficients.
As a consequence, we expect the thresholding procedure to conserve these moments, in the sense that 
\begin{equation}\label{moments_def}
\mathcal{M}_{m,k}^W = \int x_k^m f^W(\mathbf{x})\mathrm{d}\mathbf{x} \simeq \int x_k^m  f^\delta(\mathbf{x})\mathrm{d}\mathbf{x}  = \mathcal{M}_{m,k}^\delta 
\end{equation}
for $0 \leq m \leq M-1$ and for all $i\in \{1,\ldots,d\}$. 
This conservation holds up to round-off error if the wavelet coefficients can be computed exactly. 
Due to the type of wavelets that we have used, we were not able to achieve this in the results presented here. 
There remains a small error related to stages 1 and 2 of the algorithm, namely the construction of $f^H$ and the approximation of the scaling function coefficients by Eq.~(\ref{scaling_function_approximation}).
They are both of order $N_g^{-1}$.
We will present numerical examples of the moments of $f^W$ in the next section.
%It is possible to replace (\ref{scaling_function_approximation}) by the exact expression if one uses biorthogonal spline wavelets, but we have not yet considered this approach.

Conservation of moments is closely related to a peculiarity of the denoised distribution function resulting from the WBDE algorithm:
it is not necessarily everywhere positive.
Indeed, wavelets are oscillating functions by definition, and removing wavelet coefficients therefore cannot preserve positivity in general.
Further studies are needed to assess if this creates numerical instabilities when $f^W$ is used in the computation of self-consistent fields.
The same issue was discussed in Ref.~\cite{Denavit1972} where a kernel with two vanishing moments was used to linearly smooth the distribution function.
The fact that this kernel is not everywhere positive was not considered harmful in this reference.
We acknowledge that it may render the resampling of new particles from $f^W$, if it is needed in the future, more difficult.
There are ways of forcing $f^W$ to be positive, for example by applying the method to $\sqrt{f}$ and then taking the square of the resulting estimate, but this implies the loss of the moment conservation, and we have not pursued in this direction. 

%%%%%%%%%%%%%%%%%%%%%%%%%%%%
%%%%%%%%%%%%%%%%%%%%%%%%%%%%

The number of arithmetic operations to perform a fast wavelet transform from scale $2^{-J}$ to scale $2^{-L}$ with the FWT in $d$ dimensions is $2S 2^{d(J-L)}$, where $S$ is the length of the wavelet filter (12 for the Daubechies filter that we are using).
The definitions of $J$ and $L$ imply that $2^{d(J-L)}$ scales like $\frac{N_p^{\frac{2}{3}}}{\log{N_p}}$.
The cost of the binning stage of order $N_{p}$, so that the total cost for computing $f^W$ is $O(N_{p})$, not larger than the cost of one time step during the simulation that produced the data.
%To give an idea, the wavelet transform of $1024\times1024$ array under Matlab on a standard personal computer takes of the order of 1 second. 
The amount of memory needed to store the wavelet coefficients during the denoising procedure is proportional to $N_g^d$, which should at least scale like $2^{dJ}$, and therefore also like $N_p$.
If one wishes to use a finer grid to ensure high accuracy conservation of moments, the storage  requirements grow like $N_g^d$.
Thanks to optimized in-place algorithms, the amount of additional memory needed during the computation does not exceed $3S$.
Another consequence of using the FWT algorithm is that $N_g$ must be an integer multiple of $2^{J-L}$.
For comparison purposes, let us recall that most algorithms to compute the POD have a complexity proportional to $N_g^3$ when $d=2$.%  FIXME

\begin{figure}
\includegraphics[width=0.47\columnwidth]{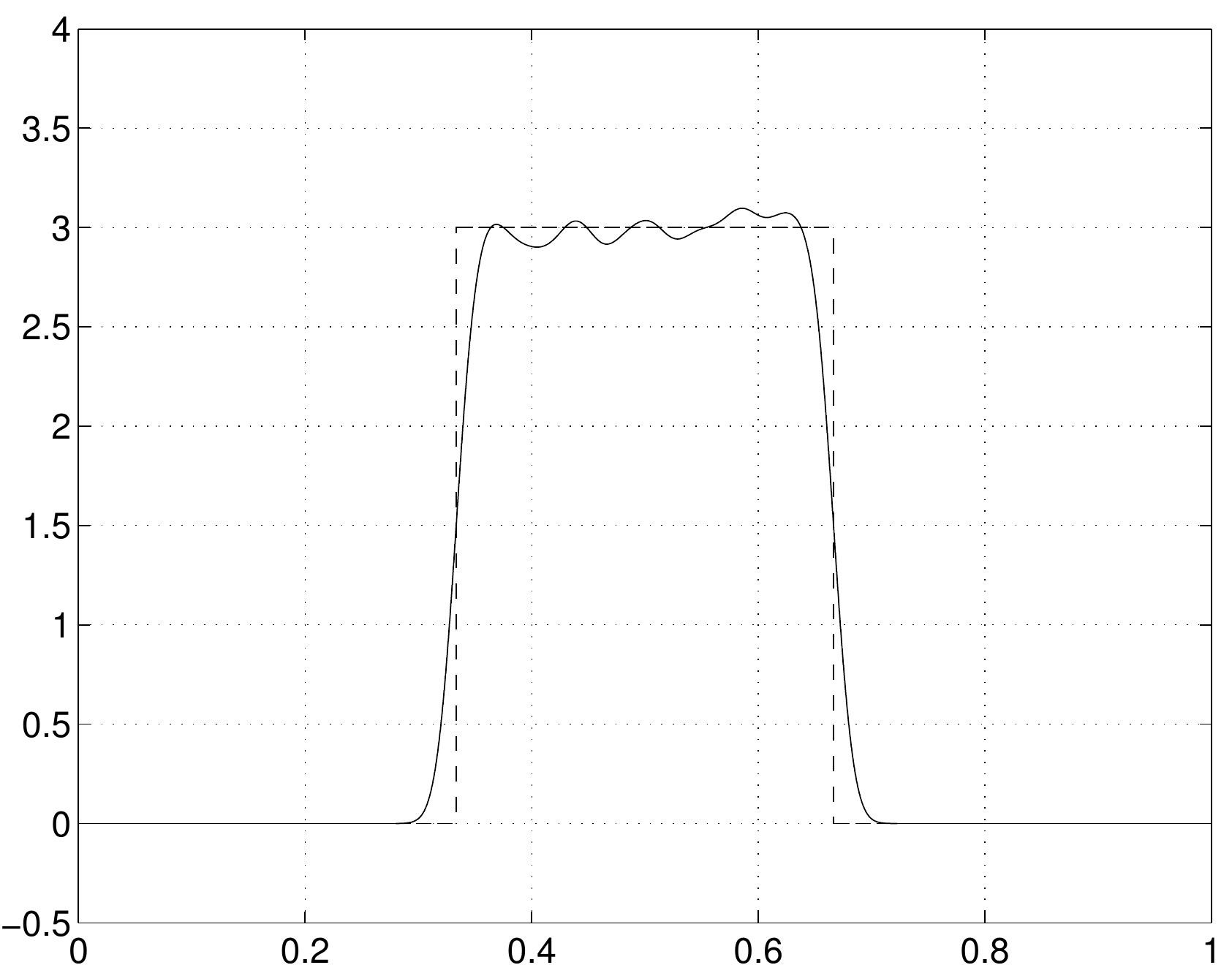}
\hspace{0.05cm}
\includegraphics[width=0.47\columnwidth]{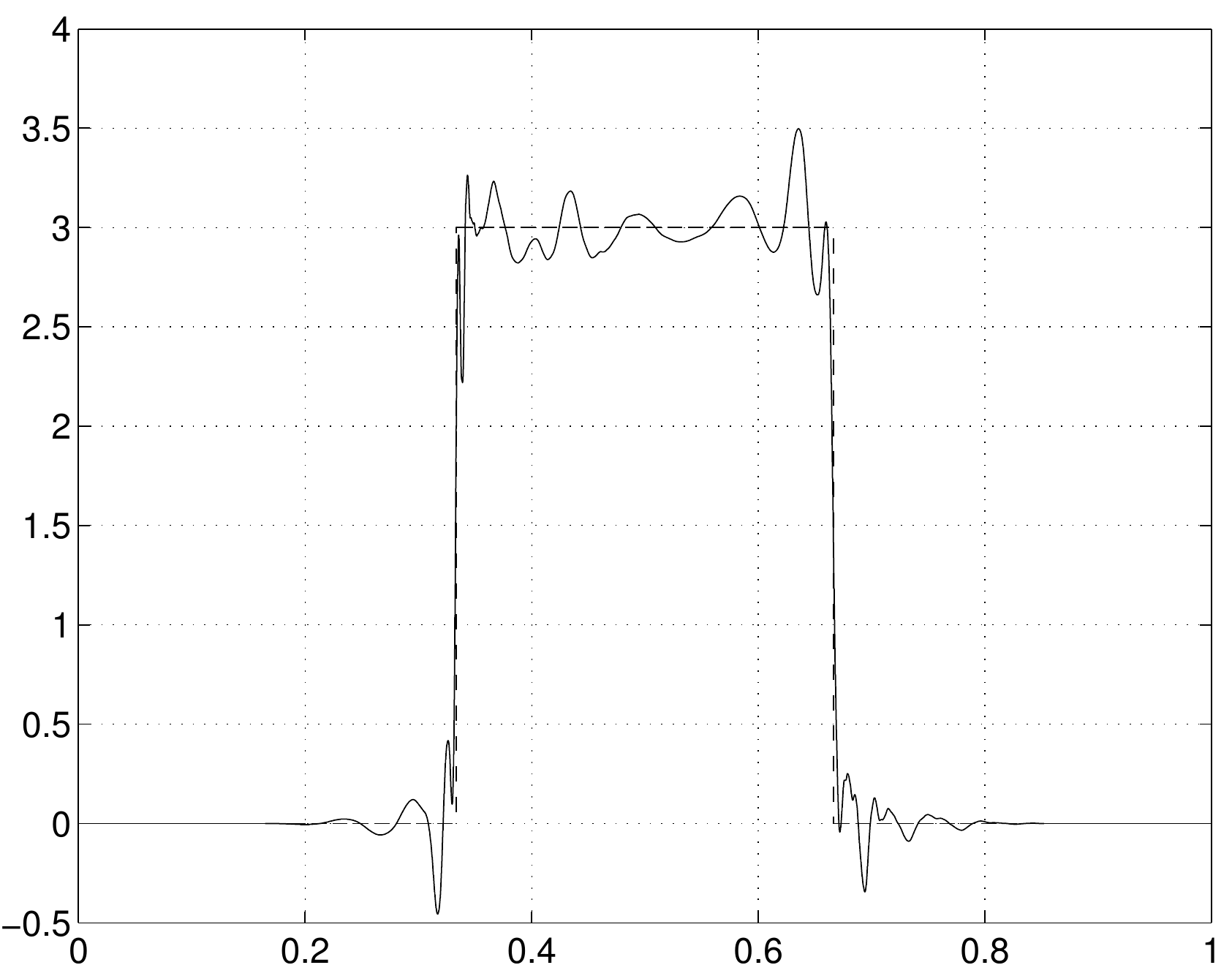}
\caption{
\label{example_1D}
Estimation of the density of a sample of size $2^{14}$ drawn uniformly in $[1/3,2/3]$,
using Gaussian kernels (left) or wavelets (right).
The discontinuous analytical density is plotted with a dashed line in the two cases.
}
\end{figure}

To conclude this subsection, Fig.~\ref{example_1D} presents an example of the reconstruction of a 1D discontinuous density that  illustrates the difference between the KDE and WBDE methods. 
The probability density function is uniform on the interval $\left[\frac{1}{3},\frac{2}{3}\right]$ and  the estimates were computed  on $\left[0,1\right]$ to include the discontinuities. 
The sample size was $2^{14}$, and the binning used $N_g = 2^{16}$ cells to compute the scaling function coefficients.
For this 1D case the value $C=2$ was used to determine the thresholds (step \ref{thresholding_function} of the algorithm).
The KDE estimate is computed using a Gaussian kernel with smoothing scale $h=0.0138$ \cite{KDE2003}.
The relative mean squared errors associated with the KDE and WBDE estimates are respectively $19.6\times 10^{-3}$ and $6.97\times 10^{-3}$.
The error in the KDE estimate comes mostly from the smoothing of the discontinuities.
The better performance of WBDE stems from the much sharper representation of these discontinuities.
It is also observed that the WBDE estimate is not everywhere positive.
The approximate conservation of moments is demonstrated on Table \ref{example_1D_moments}.
Note that the error on all these moments for $f^W$ could be made arbitrary low by increasing $N_g$.
The overshoots could also be mitigated by using nearly shift invariant wavelets \cite{NK01}.

%%%%%%%%%%%%%%%%%%%%%%%%%%%%
%%%%%%%%%%%%%%%%%%%%%%%%%%%%

%%%%%%%%%%%%%%%%%%%%%%%%%
\subsection{Proper Orthogonal Decomposition Method}
%%%%%%%%%%%%%%%%%%%%%%%%%

For completeness, in this subsection we present a brief review of the POD density reconstruction method. For the sake of comparison with the WBDE method, we limit attention to the time independent case. Further details, including the reconstruction of time dependent densities using POD methods can be found in  Ref.~\cite{delCastillo2008}.

The first step in the POD method is to construct  the histogram $f^H$ from the particle data. 
This density is represented by an $N_x \times N_y$ matrix $\hat{f}_{ij}$ containing the fraction of particles
with coordinates $(x,y)$ such that $X_i \leq x< X_{i+1}$ and $Y_i \leq y< Y_{i+1}$.  
In two dimensions, the POD method is based on the singular value decomposition of the histogram.  
According to the SVD theorem \cite{golub_van_loav_1996}, the matrix $\hat{f}$ can always be factorized as
$\hat{f}= U W V^t$ where $U$ and $V$ are $N_x \times N_x$ and $N_y \times N_y$ orthogonal matrices,
$U U^t =V  V^t = I$, and $W$ is a diagonal matrix,
$W = {\rm diag} \left( w_1, w_2, \ldots w_N \right )$, such that   $w_1 
\geq w_2\geq   \ldots \geq  w_N \geq 0$.
with $N= {\rm min} (N_x,N_y)$.

In vector form, the decomposition can be expressed as 
\bq
\label{svd_vector}
\hat{f}_{ij}= \sum_{k=1}^N\, w_k \, u^{(k)}_i\, v^{(k)}_j \, ,
\eq
where the $N_x$-dimensional  vectors, $u_i^{(k)}$, and the  $N_y$-dimensional  vectors, $v_j^{(k)}$, are the orthonormal 
POD modes and  correspond to the columns of the matrices $U$ and $V$ respectively. Given the decomposition in Eq.~(\ref{svd_vector}), we define  the rank-$r$ approximation of
$\hat{f}$    as 
\bq
\label{lr_svd}
\hat{f}^{(r)}_{ij}= \sum_{k=1}^r\, w_k \, u^{(k)}_i\, v^{(k)}_j \, ,
\eq
where $1 \leq r < N$,
and define the corresponding rank-$r$ reconstruction error as
\bq
\label{nre}
e(r) =  || \hat{f}-\hat{f}^{(r)} ||^2 = \sum_{i=r+1}^N w_i^2 \, ,
\eq
where $|| A|| = \sqrt{\sum_{i j} A_{ij}^2}$ is the Frobenius norm. 
Since $\hat{f}^{(r=N)}=\hat{f}$, we define $e(N)=0$. 
%Note that, by definition, $e(0)=||\hat{f}||^2$, and $e(N)=0$. 
The key property of the POD is that the approximation in Eq.~(\ref{lr_svd}) is 
optimal in the sense that
\bq
e(r) = {\rm min} \left  \{
||\hat{f}-g||^2 \, \left | {\rm rank} (g) = r \right. \right \} \, .
\eq
That is, of all the possible rank-$r$ Cartesian product approximations of $\hat{f}$, $\hat{f}^{(r)}$ is the closest to $\hat{f}$ 
in the Frobenius norm.  

The SVD spectrum, $\{ w_k\}$, of noise free coherent signals decays very rapidly after a few modes, but the spectrum of noise dominated signals is relatively flat and decays very slowly. 
When a coherent signal is contaminated with low  level  noise,  the SVD spectrum 
exhibits an initial rapid decay followed by a weakly decaying spectrum known as the noisy plateau.
In the POD method  the denoised density is defined as the truncation $f^P=\hat{f}^{(r_c)}$, where
$r_c$ corresponds to  the rank where the noisy plateau starts. In general it is difficult to provide a precise a priori estimate of $r_c$, and this is one of the potential limitations of the POD method. One possible  quantitative criterion used in Ref.~\cite{delCastillo2008} is to consider the relative decay of the spectrum, $\Delta(k)=(w_{k+1}-w_k)/(w_{2}-w_1)$, for $k>1$, and define $r_c$ by the condition  $\Delta(r_c)\leq \Delta_c$ where $\Delta_c$ is a predetermined threshold. 

%%%%%%%%%%%%%%%%%%%%%%%%%%%%%%%%%%%%
\begin{table}
\begin{tabular}{l*{6}{c}r}
                          & $m=0$	 	& $m=1$ 	      & $m=2$ 		    & $m=4$ \\
\hline
$f^K$               	  & $1.81\cdot 10^{-5}$	& $1.70\cdot 10^{-5}$ & $7.52\cdot 10^{-4}$ & $3.90\cdot 10^{-3}$ \\
$f^W$ 	  	  	  & $1.08\cdot 10^{-11}$ & $1.52\cdot 10^{-5}$ & $2.93\cdot 10^{-5}$ & $5.52\cdot 10^{-5}$ \\
\end{tabular}
\caption{
\label{example_1D_moments}
Relative absolute difference between the moments of $f^\delta$ and those of $f^K$ and $f^W$, for the distribution function corresponding to Fig.~\ref{example_1D}.
}
\end{table}
%%%%%%%%%%%%%%%%%%%%%%%%%%%%%%%%%%%%

%%%%%%%%%%%%%%%%%%%%%%%%%
\section{Applications}
%%%%%%%%%%%%%%%%%%%%%%%%%

In this section, we apply the WBDE method to reconstruct and denoise the particle distribution function starting from discrete particle data. The data corresponds to three  different groups of simulations: collisional thermalization with
a background plasma, guiding center transport in toroidal geometry, and Vlasov-Poisson electrostatic instabilities. The first two groups of simulations were analyzed using POD methods in Ref.~\cite{delCastillo2008}. One of the goals of this section is to compare the POD method with the WBDE method in these cases and in a new Vlasov-Poisson data set.  This data set allows the testing of the reconstruction algorithms in a collisionless system that incorporates the self-consistent evaluation of the forces acting on the particles, as opposed to the collisional, test particle problems analyzed before. 
When comparing the two methods it is important to keep in mind that POD
has one free parameter, namely
the number $r$ of singular vectors that are retained to reconstruct the denoised distribution function.
In the cases studied here we used a best guess for $r$ based on the properties of the reconstruction. In Ref.~\cite{delCastillo2008} the POD method was developed and applied to time independent and time dependent data sets. However, in the comparison with the WBDE method, we limit attention to $2$-dimensional time independent data sets. 

The accuracy of the reconstruction of the density at a fixed time $t$ will be monitored using the absolute mean square error
\begin{equation}\label{error_e}
e = \sum_{i,j} \vert f^{est}(x_i,y_j;t) - f^{ref}(x_i,y_j;t) \vert^2 \, ,
\end{equation} 
where $(x_i,y_j)$ are the coordinates of the nodes of a prescribed $N_g \times N_g$ grid in the space, and  $f^{est}$ denotes the estimated density computed from a sample with $N_p$ particles.
For the WBDE method
$f^{est}=f^W$, and for the POD method $f^{est}=f^P$. 
In principle, the reference density, $f^{ref}$, in Eq.~(\ref{error_e}) should be the density function obtained from the exact solution of the corresponding continuum model, e.g. the Fokker-Planck or the Vlasov-Poisson system. However, when no explicit solution is available, we will set $f^{ref}=f^{H}$ where  $f^{H}$ is the histogram corresponding to a simulation with a maximum number of particles available which in the cases reported here correspond to $N_p=10^6$.
We will also use the normalized error 
\bq
\label{error_e0}
e_0=\frac{e}{\sum_{i,j} \vert f^{ref}(x_i,y_j;t) \vert^2 \,} \, .
\eq

%%%%%%%%%%%%%%%%%%%%%%%%%
\subsection{Collisional thermalization with
a background plasma}
%%%%%%%%%%%%%%%%%%%%%%%%%

This first example models the relaxation of a non equilibrium plasma by collisional damping and pitch angle scattering on a thermal background.  The plasma is  spatially homogeneous and is represented by an ensemble of $N_p$ particles in a three-dimensional velocity space.  Assuming a strong magnetic field, the dynamics can be reduced to two degrees of freedom: the magnitude of the particle velocity, $v$, and the particle pitch, $\lambda=\cos \theta$, where $\theta$ is the angle between the particle velocity and the magnetic field. In the continuum limit the particle distribution function is governed by the Fokker-Planck equation which in the particle description corresponds to the  stochastic differential equations \bq
\label{mc_1}
d \lambda = - \lambda \nu_D\, dt - \sqrt{\nu_D \left ( 1 - \lambda^2 
\right) } \, d \eta_\lambda \, ,
\eq
\bq
\label{mc_2}
d v = -\left[  \alpha  \, \nu_s \, v - \frac{1}{2 v^2}\, \frac{d}{dv}\left( \nu_{||} v^4 \right) \right] \, dt + \sqrt{v^2\, \nu_{||}} \, 
d\eta_{v}\, ,
\eq
describing the evolution of $v \in (0, \infty)$ and $\lambda \in [-1,1]$ for each particle,
where $d \eta_\lambda$ and $d \eta_v$ are independent Wiener
stochastic processes and $\nu_D$, $\nu_s$ and $\nu_{\parallel}$ are functions of $v$. For further details on the  model  see  Ref.~\cite{delCastillo2008} and references therein. 

We considered simulations with $N_p=10^3$, $10^4$, $10^5$ and $10^6$ particles. The initial conditions of the ensemble of particles were obtained by sampling a distribution of the form
\bq
\label{f_ic}
f(v,\lambda ,t=0)= C v^2 \exp \left \{ -\frac{1}{2}\left[ \frac{(\lambda-\lambda_0)^2}{\sigma_\lambda^2}
+ \frac{(v-v_0)^2}{\sigma_v^2}\right] \right \} \, ,
\eq
where a $v^2$ factor has been included in the definition of the initial condition so that the volume element is simply $\mathrm{d}v \mathrm{d}\mu$, $C$ is a normalization constant, $\lambda_0=0.25$, $v_0=5$, $\sigma_\lambda=0.25$ and $\sigma_v=0.75$.  
This relatively simple problem is particularly well suited for the WBDE method because 
the simulated particles do not interact and therefore
statistical correlations can not build-up between them.

\begin{figure}%[htbp]
\includegraphics[width=0.49\columnwidth]{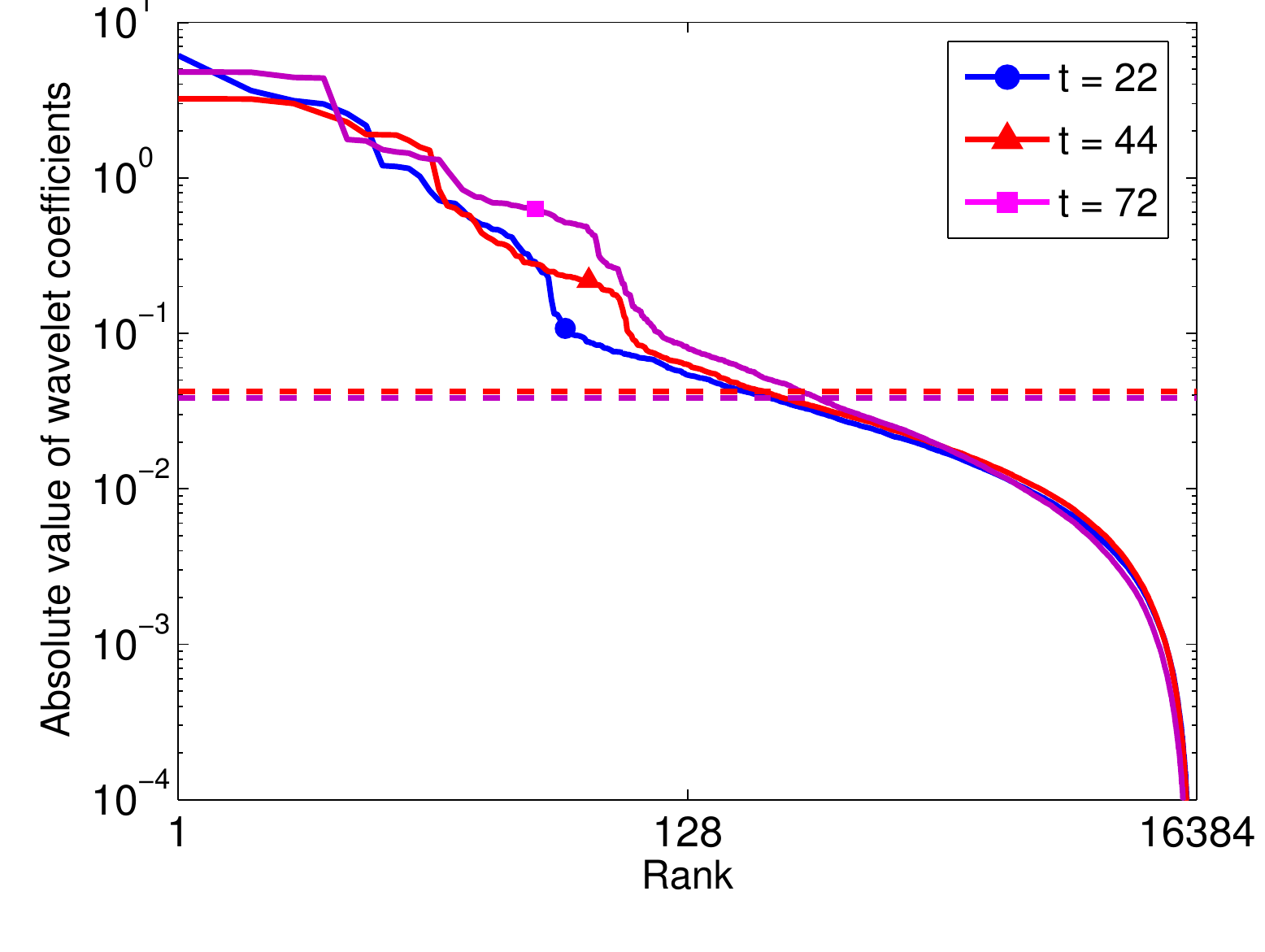}
\includegraphics[clip,width=0.49\columnwidth]{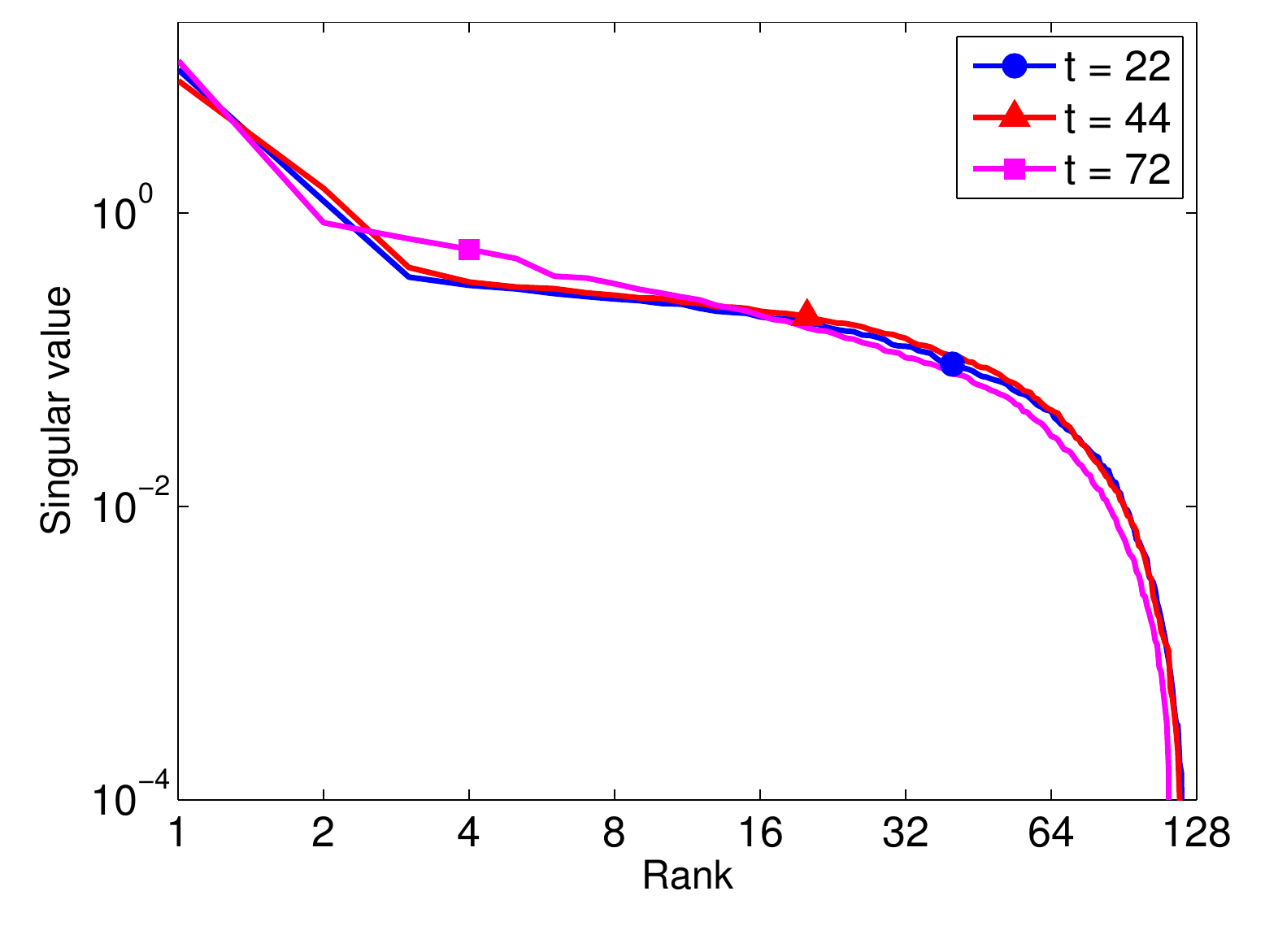}
\includegraphics[width=0.49\columnwidth]{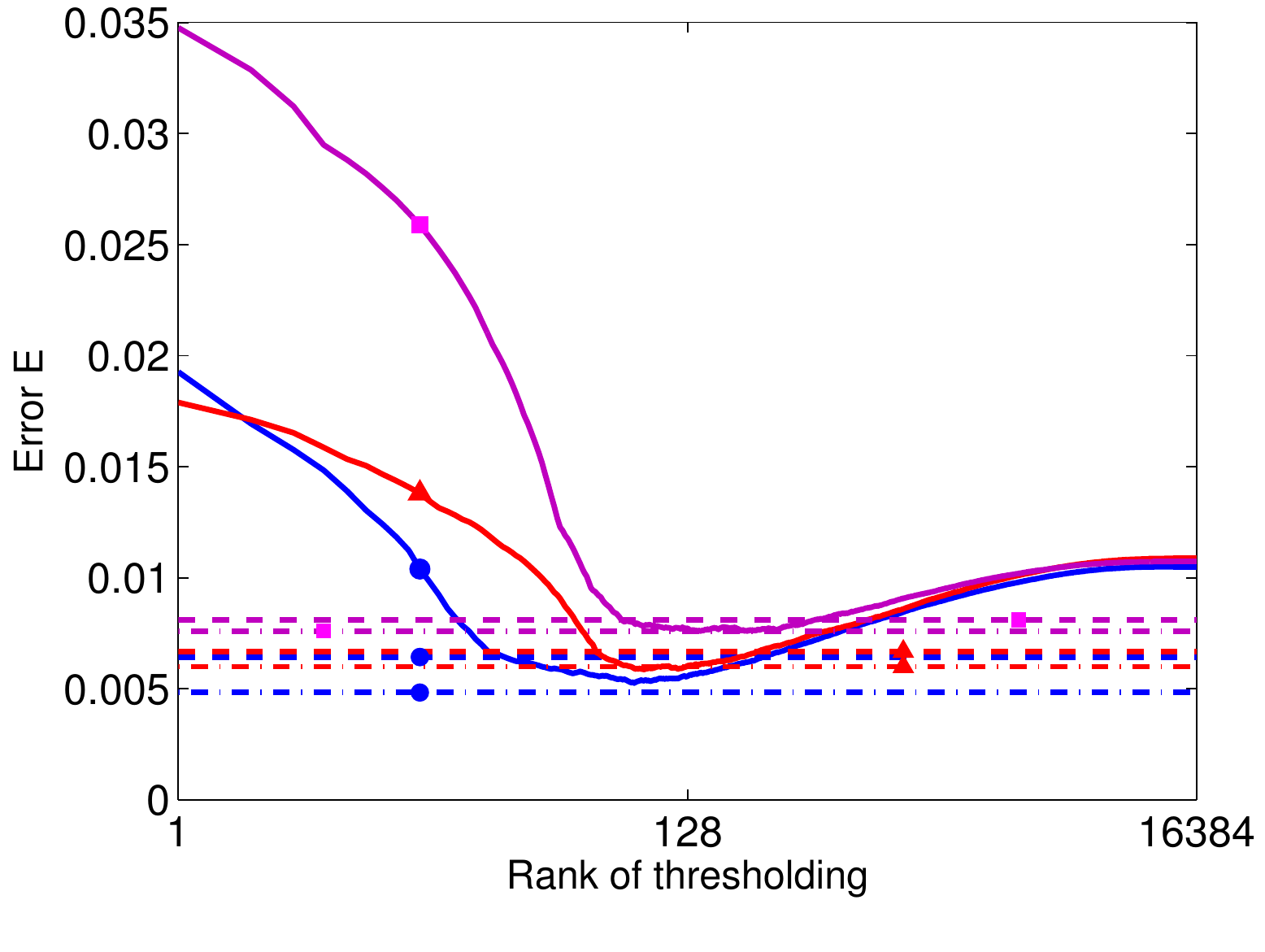}
\includegraphics[width=0.49\columnwidth]{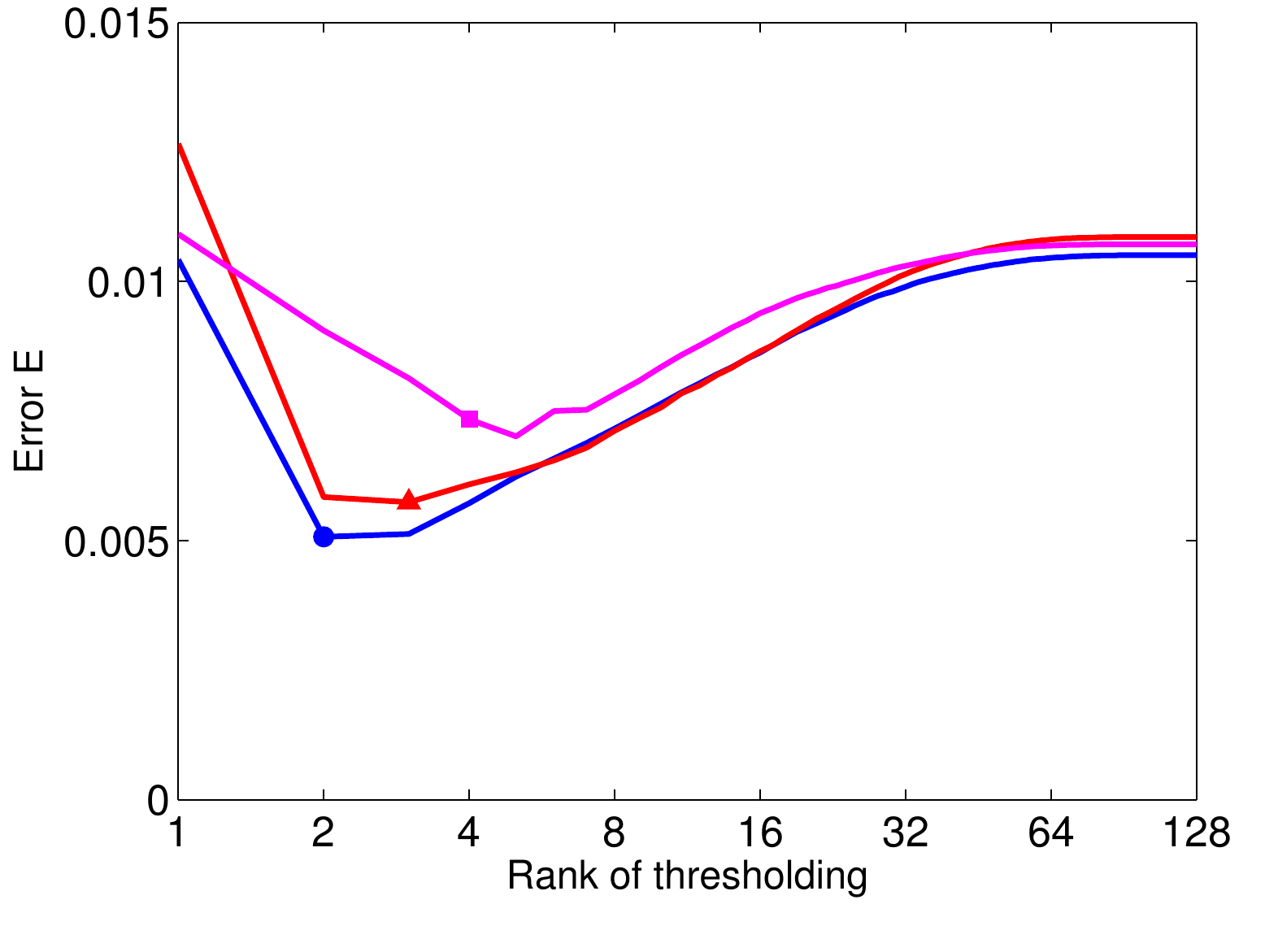}
\caption{
\label{compression_curve_time_dependent}
Wavelet and POD analyses of collisional  relaxation particle data at different fixed times, with $N_p = 10^5$.
Top left: absolute values of the wavelet coefficients sorted by decreasing order (full lines), and thresholds given by the Waveshrink algorithm (dashed lines).
Top right: singular values of the histogram used to construct $f^P$.
Bottom left: 
error estimate $\frac{e^{1/2}}{N_g^2} $ with respect to the run for $N_p = 10^6$ as a function of the number of retained wavelet coefficients (full lines), error obtained when using the Waveshrink threshold (dashed lines),
and error obtained using the WBDE method (dash-dotted lines).
Bottom right: 
error estimate $\frac{e^{1/2}}{N_g^2}$ for $f^P$ as a function of the number $l$
of retained singular values.
}
\end{figure}

Before applying the WBDE method, we analyze the sparsity of the wavelet expansion of $f^\delta$, and compare the number of modes kept and the reconstruction error for different thresholding rules.
%One way of visualizing the 
The plot in the upper left panel of Fig.~\ref{compression_curve_time_dependent} shows the absolute values of the wavelet coefficients in decreasing order at different fixed times.
The wavelet coefficients exhibit a clear rapid decay beyond the few significant modes corresponding to the gross shape of the Maxwellian distribution. A similar trend is observed in the coefficients of the POD expansion shown in 
the upper right panel of Fig.~\ref{compression_curve_time_dependent}.
However, in the wavelet case  the exponential decay starts after more than $100$ modes, whereas in the POD case the exponential decay starts after only one mode.

%%%%%%%%%%%%%%%%%%%%%%%%%%%%%%%%%%%%%
%%%%%%%%%%%%%%%%%%%%%%%%%%%%%%%%%%%%%

The two panels at the bottom of Fig.~\ref{compression_curve_time_dependent} show the square root of the reconstruction error normalized by $N_g$, $\sqrt{e}/N_g^2$,  in the WBDE and POD methods. Because in this case we do not have access to the exact solution of the corresponding Fokker-Planck equation at the prescribed time, we used $f^H$ computed using $N_p=10^6$ particles as the reference density $f^{ref}$ in Eq.~(\ref{error_e}).
The error observed when applying a global threshold to the wavelet coefficients (bottom left panel in Fig.~\ref{compression_curve_time_dependent}) is minimal when around $100$ modes are kept whereas in the POD case 
(bottom right panel in Fig.~\ref{compression_curve_time_dependent}) the minimal error is reached with about two or three modes. 
Fig.~\ref{compression_curve_time_dependent} also shows the wavelet threshold obtained by applying the iterative algorithm 
based on the stationary Gaussian white noise hypothesis \cite{AAMF04,Farge2006}.
The error corresponding to this threshold is larger than the optimal error because the noise in this problem is very non-stationary due to the lack of statistical fluctuations in the regions were particles are absent.
In contrast, the error corresponding to the WBDE procedure (dash-dotted line) is typically smaller than the optimal error obtained by global thresholding.This is not a contradiction, because the WBDE procedure is not a global threshold, but a level dependent threshold.

%%%%%%%%%%%%%%%%%%%%%%%%%%%%%%%%%%%%
\begin{figure}%[htbp]
\includegraphics[width=1.0\columnwidth]{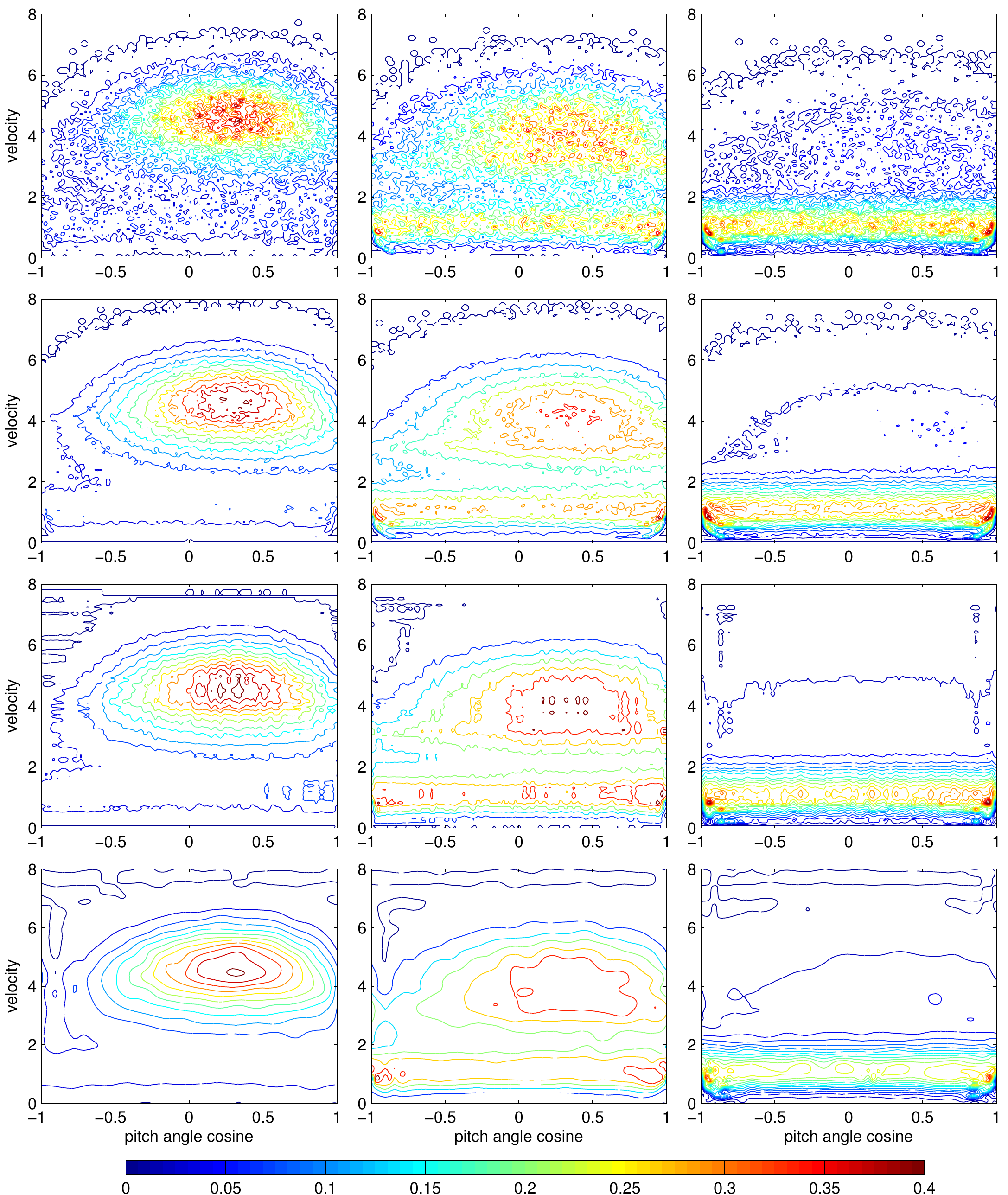}
\caption{\label{time_dependent_contour_plots}
Contour-plots of estimates of $f$ for the collisional relaxation particle data. 
First row: Histogram method estimated using $N_p=10^{5}$ particles.
Second row: Histogram method estimated using $N_p=10^{6}$ particles.
Third row: POD method estimated using $N_p=10^5$ particles.
Fourth row: WBDE method estimated using $N_p=10^5$ particles.
The three columns correspond to $t=28$, $t=44$ and $t=72$ respectively.
The plots show twenty isolines, equally spaced in the interval $[0,0.4]$.
}
\end{figure}
%%%%%%%%%%%%%%%%%%%%%%%%%%%%%%%%%%%%

Figure~\ref{time_dependent_contour_plots} compares at different times the densities estimated with the WBDE and the POD (retaining only three modes) methods using $N_p=10^5$ particles with the histograms computed using  $N_p=10^5$ and $10^6$  particles. The key  future to observe is that 
the level of smoothness of $f^W$ and $f^P$ corresponding to $N_p=10^5$ is similar, if not greater, than the level of smoothness in $f^H$ computed using ten times more particles, i.e. $N_p=10^6$ particles. 
Table~\ref{time_dependent_error} summarizes the normalized reconstruction errors for $N_p=10^5$  according  Eq.~(\ref{error_e}) using $f^H$ with $N_p=10^6$ as $f^{ref}$. 
The WBDE and POD denoising methods offer a significant improvement, approximately by a factor $2$, over the 
raw histogram method.

\begin{table}
\vspace{3 cm}
\begin{tabular}{l*{6}{c}r}
                          & $t = 28$ & $t = 44$ & $t = 72$  \\
\hline
$f^H$               	  & $0.14$ & $0.17$ & $0.12$  \\
$f^P$ 	  	  	  & $0.068$ & $0.090$ & $0.094$  \\
$f^W$ 	  	  	  & $0.064$ & $0.094$ & $0.088$  \\
% non normalized e :
% $f^H$               	  & $1.34$ & $1.39$ & $1.37$  \\
% $f^P$ 	  	  & $0.62$ & $0.77$ & $0.97$  \\
% $f^W$ 	  	  & $0.66$ & $0.74$ & $1.04$  \\
\end{tabular}
\caption{
\label{time_dependent_error}
Normalized root mean squared error $e_0$ (\ref{error_e0}) for the histogram, POD and WBDE estimates
of the particle distribution function for $N_p=10^5$ at three different times of
the Maxwellian relaxation problem.
}
\end{table}

\begin{figure}%[htbp]
\includegraphics[width=0.8\columnwidth]{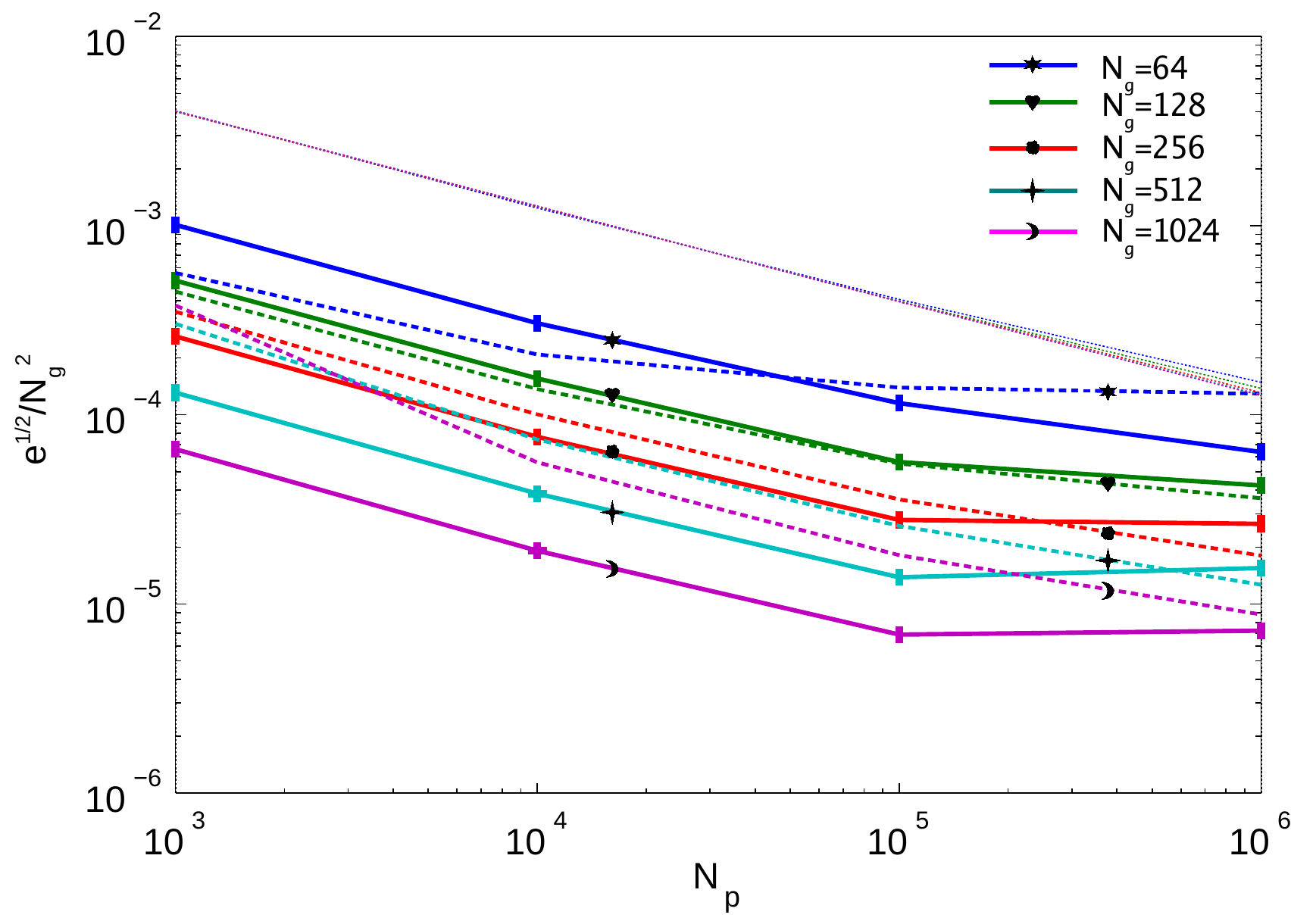}
\caption{\label{wavelets_e}
Reconstruction error, $\frac{e^{1/2}}{N_g^2} $, as a function of $N_{p}$ for the collisional relaxation particle data
corresponding to the Maxwellian equilibrium state. 
Bold solid lines correspond to the WBDE method, bold dashed lines correspond to the POD method, and thin dashed lines  
correspond to the histogram method.}
\end{figure}

A more detailed comparison of the estimates can be achieved by focusing on the Maxwellian final equilibrium state
\bq
\label{f_max}
f_M(v)= \frac{2}{\sqrt{\pi} } v^2 e^{-v^2} \, ,
\eq
where, as in Eq.~(\ref{f_ic}), the  $v^2$ metric  factor has been included in the definition of the distribution.
For this calculations we considered sets of particles sampled from Eq.~(\ref{f_max}) in the compact domain
$[-1,1]\times[0,4]$. Since $f_M$ is an exact equilibrium solution of the Fokker-Plack equation, the ensemble of particles will be in statistical equilibrium but it will exhibit fluctuations due to the finite number of particles. 
Figure~\ref{wavelets_e} shows the dependence of the square root of the reconstruction error, $e$ (normalized by $N_g^2$) on
the number of particles $N_p$ and the grid resolution $N_g$ for the WBDE and POD methods. The main advantage of this example is that the exact density $f^M$ can be used as the reference density $f^{ref}$ in the evaluation of the error.

%%%%%%%%%%%%%%%%%%%%%%%%%%%%%%%%%%%%%
%%%%%%%%%%%%%%%%%%%%%%%%%%%%%%%%%%%%

%%%%%%%%%%%%%%%%%%%%%%%%%
\subsection{Collisional guiding center transport in toroidal geometry}
\label{delta_5d_example}
%%%%%%%%%%%%%%%%%%%%%%%%%

The previous example focused on collisional dynamics. However, in addition to collisions, plasma transport involves external and self-consistent electromagnetic fields  and it is of interest to test the particle density reconstruction algorithms in these more complicated settings.  As  a first step on this challenging problem we consider a plasma subject to collisions and an externally applied fixed magnetic field in toroidal geometry. 
The choice of the field geometry and structure was motivated by problems of interest to magnetically confined fusion plasmas. The data  was presented and analyzed using POD method in 
Ref.~\cite{delCastillo2008}. 
The phase space of the simulation is five dimensional. However,  as in Ref.~\cite{delCastillo2008}, we limit attention to the denoising of the particles distribution function along two coordinates corresponding to the poloidal angle $\theta\in[0,2\pi]$ and the cosine of the pitch angle $\mu\in[-1,1]$.
The remaining three coordinates have been averaged out for the purpose of this study.
The $\theta$ coordinate is periodic, but the pitch coordinate $\mu$ is not. 

An important issue to consider is that the data was generated using a $\delta f$ code (DELTA5D). Based on an expansion on $\rho/L \ll1$ (where $\rho$ is the characteristic Larmor radius and $L$ a typical equilibrium length scale) the distribution function is decomposed into a Maxwellian part $f_M$ and a first-order perturbation $\delta f$ represented as a collection of particles (markers)
\bq
\delta f({\bf x}) = \sum_n W_n \delta({\bf x}-{\bf X}_n) \, ,
\eq
like in Eq.~(\ref{dirac_estimate}) except that each marker is assigned a time dependent weight $W_n$ whose time evolution depends on the Maxwellian background \cite{parker}. 
The direct use of $\delta f({\bf x})$ is problematic in the WBDE method because $\delta f$ is not a probability density. To circumvent this problem the WBDE method was applied after normalizing the $\delta_f$ distribution so that
$\int \vert\delta f\vert^H =1$, on a  $128\times128$ grid. 

\begin{figure}%[htbp]
\includegraphics[width=1.0\columnwidth]{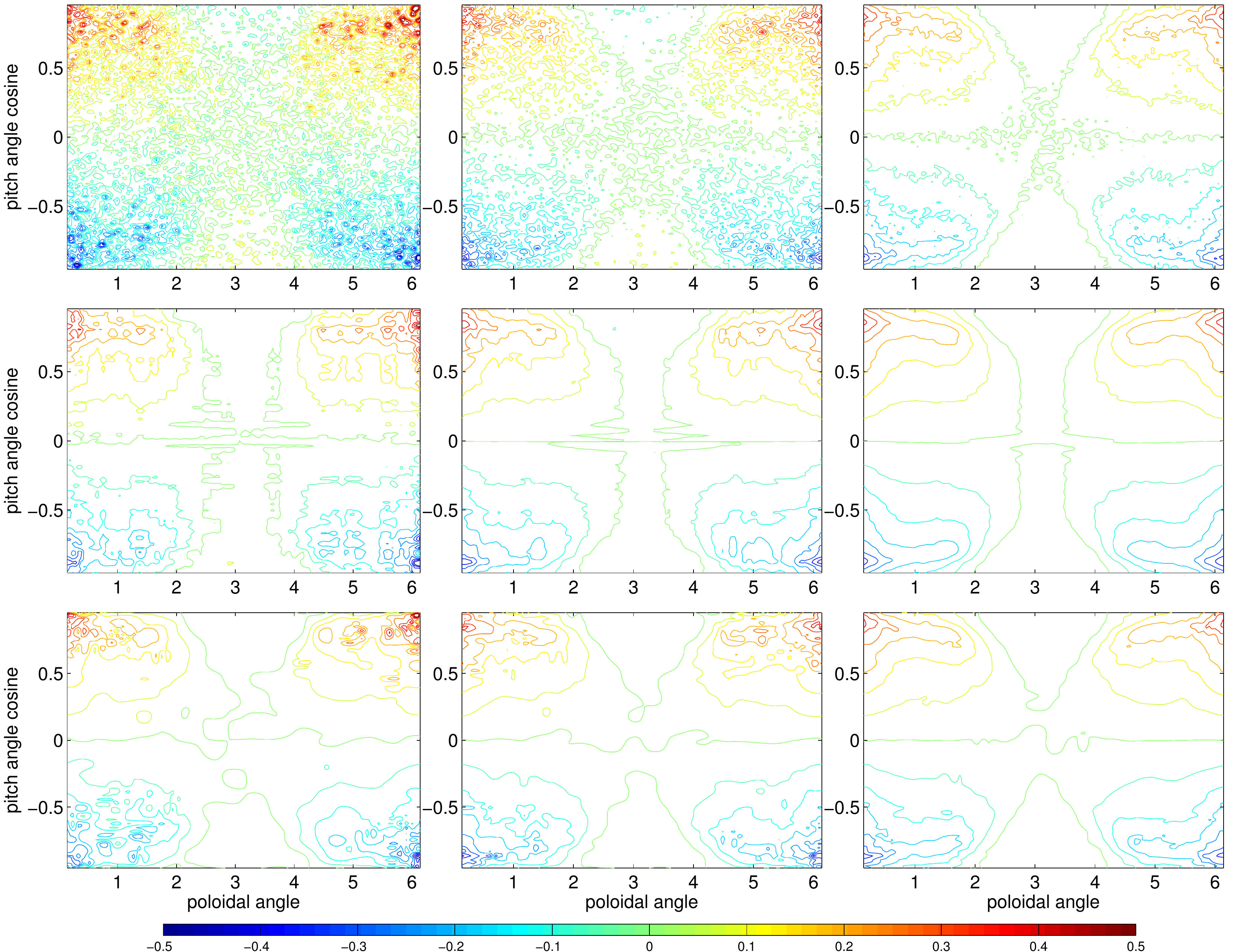}
\caption
{
\label{fig:d5d_3d_comp}
Contour plots of estimates of $f$ for the collisional guiding center transport particle data:
Histogram method  (first row), POD method (second row), and WBDE method (third row).
The left, center and right 
columns correspond to  $N_p = 32\cdot 10^3$ (left), $N_p = 128\cdot 10^3$ (middle) and $N_p = 1024\cdot 10^3$ (right)
respectively. 
The plots show seventeen isolines equally spaced within the interval $[-0.5,0.5]$.
}
\end{figure}

\begin{figure}%[htbp]
%\begin{centering}
{\includegraphics[width=0.8\columnwidth]
{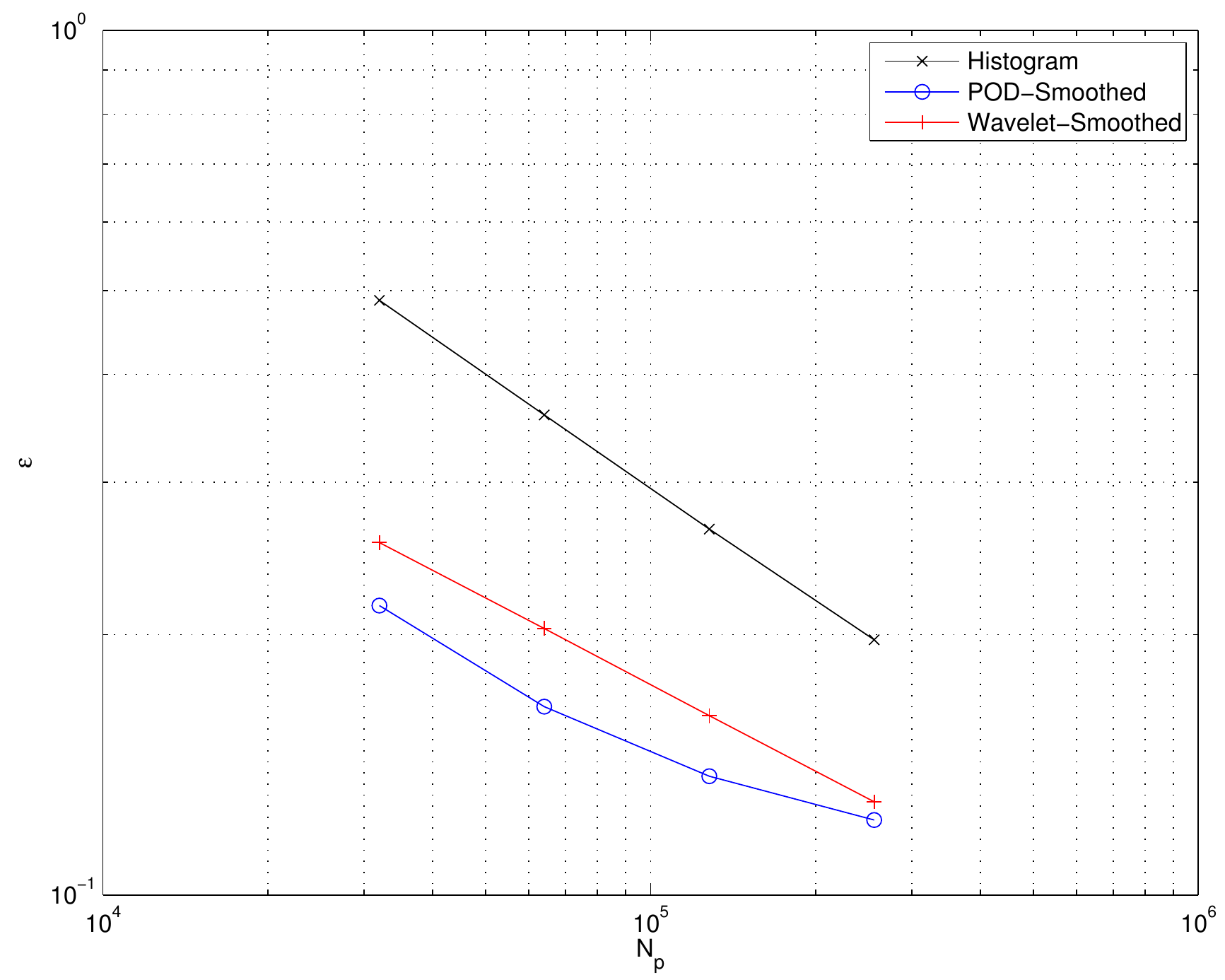}
}
%\end{centering}
\caption
{
\label{fig:d5d_RMS}
Error estimate, $\frac{e^{1/2}}{N_g^2} $, for collisional guiding center transport particle data
according to the histogram, the POD, and the wavelet methods. 
}
\end{figure}

Figure~\ref{fig:d5d_3d_comp} shows contour plots of the histogram $f^H$ corresponding to $N_p=32 \times 10^3$, $64 \times 10^3$, and $1024 \times 10^3$ along with the WBDE and POD reconstructed densities. The POD reconstructions were done using $r=3$ modes, as in Ref.~\cite{delCastillo2008}.
It is observed that comparatively high levels of smoothness can be achieved with considerably less particles by using either the  WBDE or  POD reconstruction methods. 
The WBDE method provides better results for the $\delta f \sim 0$ contours. This is because
POD modes are tensor product functions, that have difficulties in approximating the triangular shape of these contour lines.
Note that the boundary artifacts due to periodization of the Daubechies wavelets do not seem to be very critical. The large wavelet coefficients associated with the discontinuity between the values of $\delta f$ at $\mu=\pm1$ are not thresholded, 
so that the discontinuity is preserved in the denoised function. 
Figure~\ref{fig:d5d_RMS} compares the reconstruction errors in the  WBDE, POD, and histogram methods as functions of the number of particles. To evaluate the error we used $f^H$ computed using $N_p=1024 \times 10^3$ as the reference density $f^{ref}$.
As in the collisional transport problem, the error is reduced roughly by a factor $2$ for both methods compared to the raw histogram. Note that the scaling with $N_p$ is slightly better for WBDE than for POD.

% %%%%%%%%%%%%%%%%%%%%%%%%%%%%%%%%%%%%%
% %%%%%%%%%%%%%%%%%%%%%%%%%%%%%%%%%%%%%

% %%%%%%%%%%%%%%%%%%%%%%%%%%%%%%%%%%%%%
% %%%%%%%%%%%%%%%%%%%%%%%%%%%%%%%%%%%%%

%%%%%%%%%%%%%%%%%%%%%%%%%
\subsection{Collisionless electrostatic instabilities}\label{PIC_example}
%%%%%%%%%%%%%%%%%%%%%%%%%

In this section we apply the WBDE and POD methods to reconstruct the single particle distribution function from discrete particle data obtained from PIC simulations of a Vlasov-Poisson plasma.
We consider a one-dimensional, electrostatic, collisionless electron plasma with an ion neutralizing background in a finite size domain with periodic boundary conditions.  
In the continuum limit the  dynamics  of the distribution function is governed by the system of equations
\begin{eqnarray}
\partial_t f + v \partial_x f + \partial_x \phi \partial_v f=0
\\
\partial_x^2 \phi = \zeta \int f(x,v,t) dv -1 \, ,
\end{eqnarray}
where the  variables have been non-dimensionalized using the Debye length as length scale  and  the plasma frequency as time scale, and $L$ is the length of the system normalized with the Debye length. 
Following the standard PIC methodology \cite{Birdsall1985}, we solve the Poisson equation on a grid and solve the particle equations using a leap-frog method.
The reconstruction of the charge density uses a triangular shape function.
We consider two initial conditions:
the first one leads to a bump on tail instability, and the second one to a two streams instability.

%%%%%%%%%%%%%%%%%%%%%%%%%
\subsubsection {Bump on tail instability}\label{bump_tail}
%%%%%%%%%%%%%%%%%%%%%%%%%

\begin{figure}%[htbp]
\includegraphics[width=0.9\columnwidth]{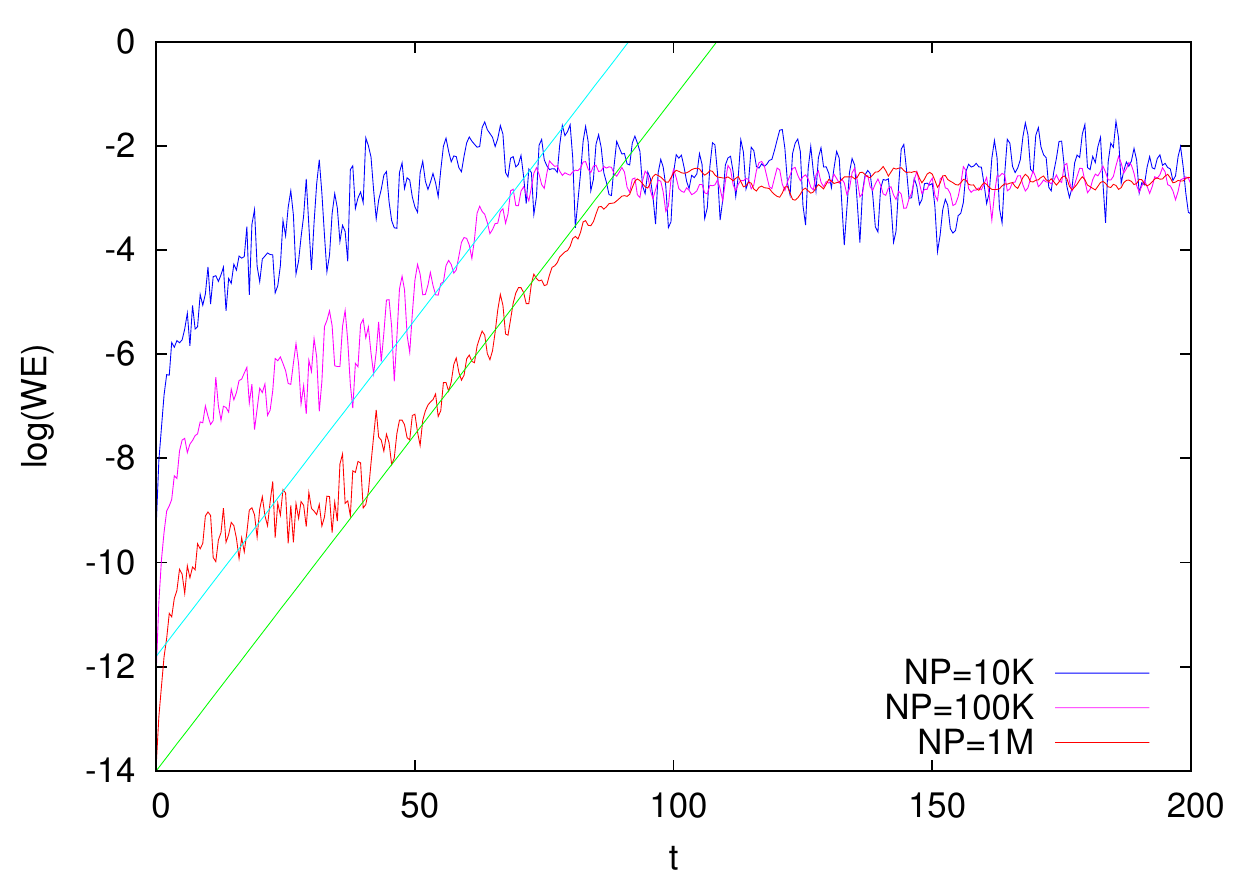}
\caption
{
\label{pic_validation}
Electrostatic energy as a function of time in the Vlasov-Poisson PIC simulations of the bump on tail instability for different numbers of particles.
The straight lines denote the growth rate predicted by linear stability theory \cite{delCastillo1998}.
}
\end{figure}

For the bump on tail instability we initialized ensembles of particles by sampling the distribution function
\begin{equation}
\label{bump_on_tail_init}
f_0(x,v) = \frac{2}{3 \pi \zeta}\frac{1-2 q v+ 2 v^2}{\left( 1 + v^2 \right)^2}\, .
\end{equation}
using a pseudo-random number generator. 
This equilibrium is stable for $q \leq 1$ and unstable for $q>1$.
The dispersion relation and linear stability analysis for this equilibrium studied in Ref.~\cite{delCastillo1998} was used to benchmark the PIC code as shown in Fig.~\ref{pic_validation}.
In all the computations presented here $q=1.25$ and $N_p=10^4$, $10^5$ and $10^6$.
The spatial domain size was set to $\zeta=16.52$  to fit the wavelength of the most unstable mode. 

%%%%%%%%%%%%%%%%%%%%%%%%%%%%%%%%%%%%
%%%%%%%%%%%%%%%%%%%%%%%%%%%%%%%%%%%%

Since the value of $q$ is relatively close to the marginal value, the instability grows weakly 
and is concentrated in a narrow band in phase space centered around the point where the bump is located, 
$v \approx 1$ in this case.
In order to unveil the nontrivial dynamics we focus the analysis in the band $v\in(-3,3)$, 
and plot  the departure of the particle distribution function from the initial background equilibrium. 
The POD method is applied directly to $\delta f^H=f^H(x,v,t)-f_0(x,v)$,
but the WBDE method is applied to the full $f^H(x,v,t)$, and $f_0(x,v)$ is subtracted only for visualization.
Note that because we are considering only a subset of phase space, 
the effective numbers of particles, $N_p=7318$, $N_p=73143$ and $N_p=731472$,
are smaller than the nominal numbers of particles, $N_p=10^4$, $N_p=10^5$ and $N_p=10^6$ respectively.

\begin{figure}%[htbp]
\includegraphics[width=1.0\columnwidth]{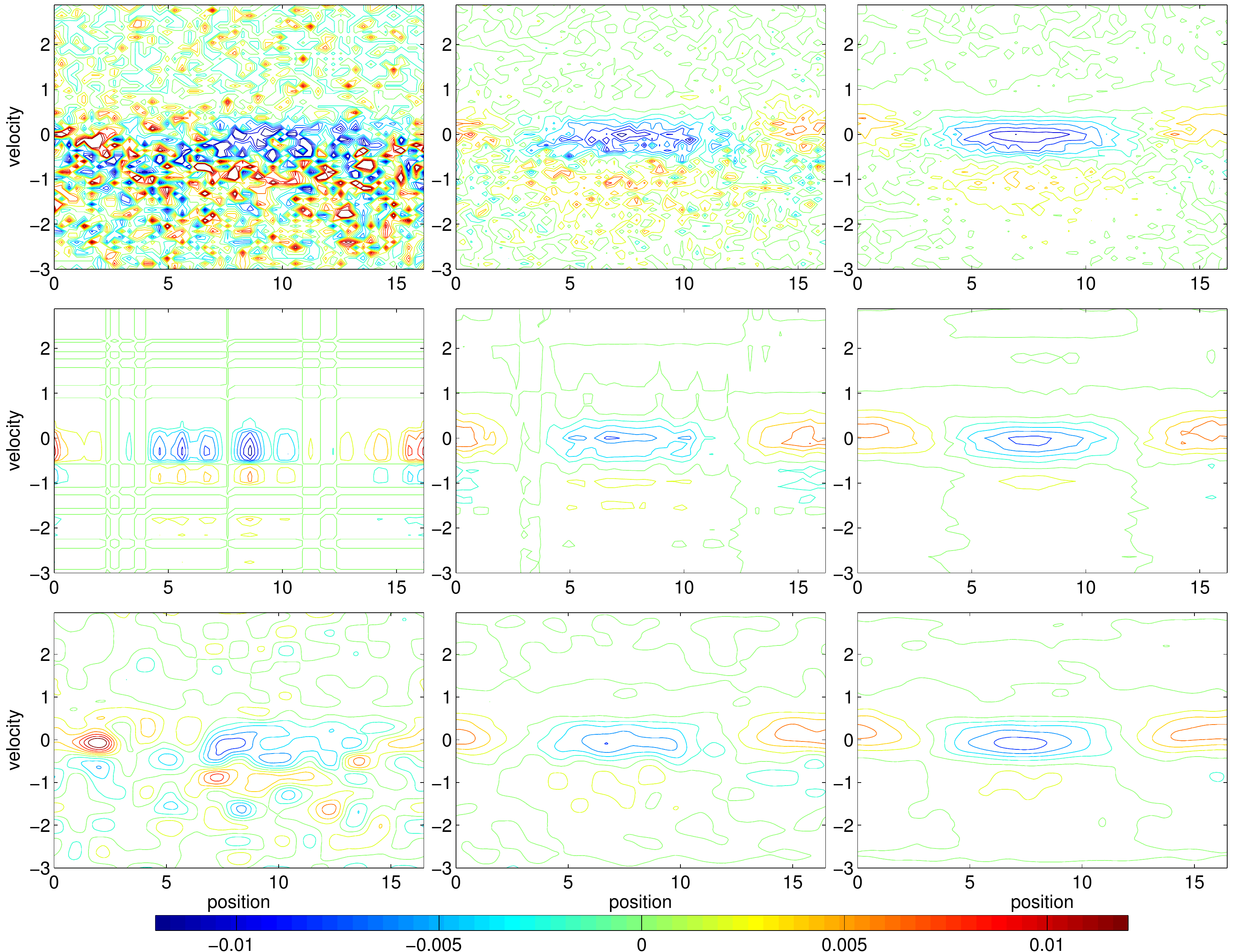}
\caption
{
\label{bump_tail_snapshots}
Contour plots of estimates of $\delta f$ for the bump-on-tail instability PIC data at $t=149$:
Histogram method  (first row), POD method (second row), and WBDE method (third row),
The left, center and right 
columns correspond to $N_p=10^4$, $N_p=10^5$ and $N_p=10^6$ particles respectively.
The plots show thirteen contour lines equally spaced within the interval $[-0.012 0.012]$.
}
\end{figure}

Figure~\ref{bump_tail_snapshots} shows contour plots of $\delta f$, for different number of particles. 
Since the instability is seeded only by the random fluctuations in the initial condition, increasing  $N_p$ delays the onset of the linear stability and this leads to a phase shift of the nonlinear saturated regime. 
To aid the comparison of the saturated regime for different numbers of particles we have eliminated this phase shift by centering the peak of the particle distributions  in the middle of the computational domain. 
A $256 \times 256$ grid was used in the  WBDE method, and a  $50\times 50$  grid was used for the histogram and the  POD methods.
The thresholds for the POD method where  $r=1$, $r=2$, and $r=3$  for $N_p=10^4$, $N_p=10^5$ and $N_p=10^6$, respectively.
Except for the case where $N_p=10^4$, both the POD and WBDE estimates are very smooth, in agreement with the expected behavior of $f$ for this instability.
It is observed that the level of smoothness of the histogram estimated using $10^6$ particles is comparable to the level of smoothness achieved after denoising using only $10^5$ particles.
One should mention that for scales between $L$ and $J$ occurring in 
the WBDE algorithm we find that none of the wavelet coefficients are above the thresholds at each scale. 
In fact, a simple KDE estimate with a large enough smoothing scale would probably do the job pretty well for this kind of instabilities which do not induce abrupt variations in $f$.
Table~\ref{bump_tail_error} shows the POD and WBDE reconstruction errors for $N_p=10^4$ and $N_p=10^5$.
The error is computed using formula (\ref{error_e0}), taking for $f_{ref}$ the histogram obtained from the simulation with $N_p=10^6$.

%%%%%%%%%%%%%%%%%%%%%%%%%%%%%%%%%%%%
%%%%%%%%%%%%%%%%%%%%%%%%%%%%%%%%%%%%

\begin{figure}%[htbp]
\includegraphics[width=0.8\columnwidth]{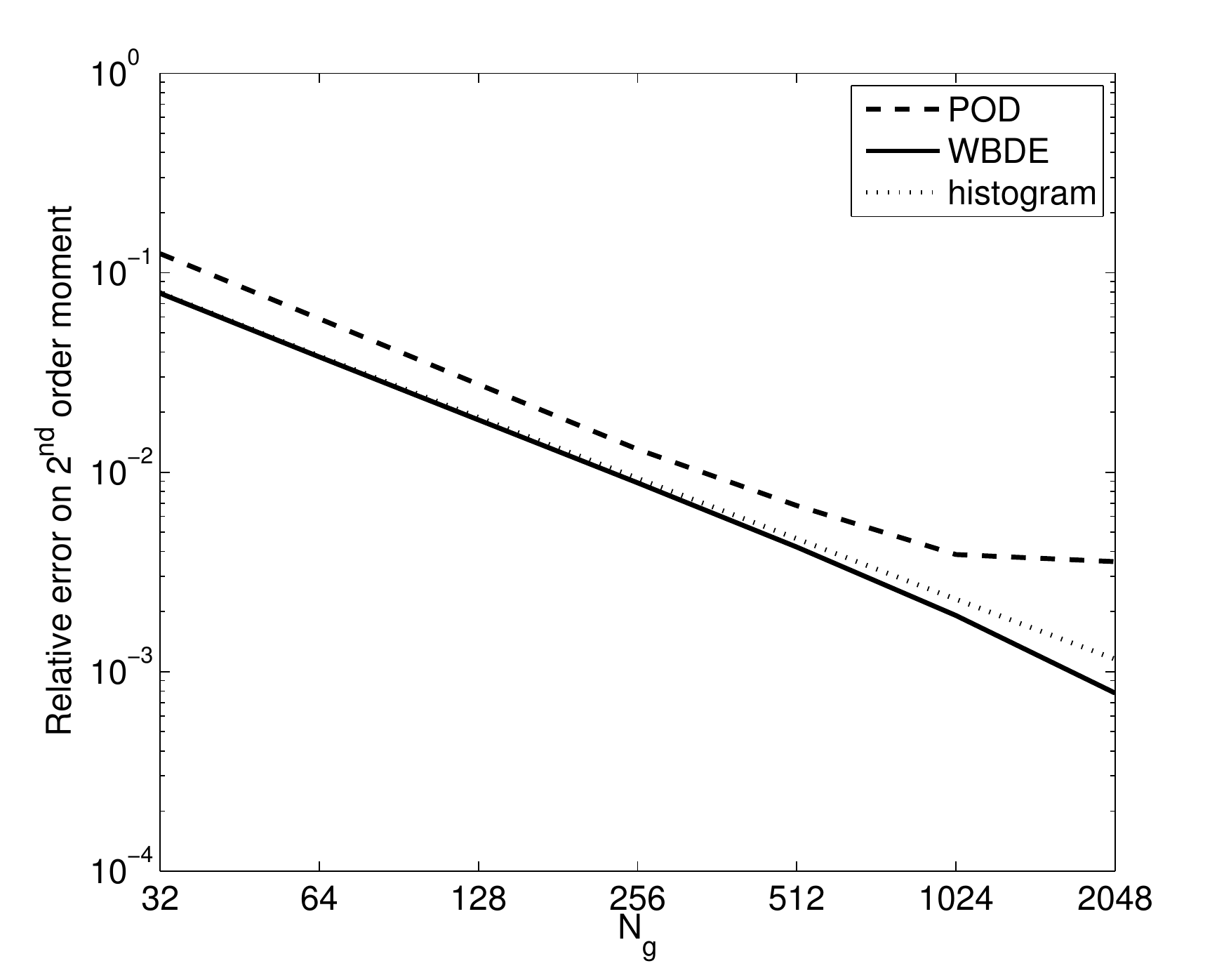}
\caption
{
\label{pic_moments}
Relative error on the second order moment as a function of the grid resolution, $N_g$,  in the POD, WBDE, and histogram methods for the bump on tail instability particle data at $t=149$, with $N_p=10^6$ particles.
}
\end{figure}

Figure~\ref{pic_moments} shows the relative error on the second order moment : 
$$\frac{\vert \mathcal{M}_{v,2}^W - \mathcal{M}_{v,2}^\delta \vert}{ \mathcal{M}_{v,2}^\delta }$$ 
where $\mathcal{M}^W_{v,2}$ is defined by (\ref{moments_def}).
A similar quantity is also represented for $f^H$ and $f^P$.
The time and number of particles are kept fixed at $t=149$ and $N_p=10^6$, and the grid resolution is varied.
As expected, $f^H$ and $f^W$ conserve the second order moment with accuracy $O(N_g^{-1})$.
The errors corresponding to $f^P$ is of the same order of magnitude but seems to reach a plateau for $N_g \simeq 1024$.
This may be due to the fact that for $N_g \geq 1024$, there is less than one particle per cell of the histogram used to compute $f_P$.

%%%%%%%%%%%%%%%%%%%%%%%%%%%%%%%%%%%%
%%%%%%%%%%%%%%%%%%%%%%%%%%%%%%%%%%%%

%%%%%%%%%%%%%%%%%%%%%%%%%
\subsubsection {Two-streams instability}\label{two_streams}
%%%%%%%%%%%%%%%%%%%%%%%%%

As a second example we consider the standard two-streams instability with an initial condition consisting of two counter-propagating cold electron beams initially located at $v=-1$ and $v=1$. 
This case is conceptually different to the previous one 
because the initial condition depends trivially on the velocity.  Therefore,  there is no statistical error in the sampling of the distribution and the noise builds up only due to the self-consistent interactions between particles.
In other words, there is initially a strong correlation between particles' coordinates, which will eventually
almost vanish. This situation offers a  way to test robustness of the WBDE method with respect to the underlying decorrelation hypothesis.

%%%%%%%%%%%%%%%%%%%%%%%%%%%%%%%%%%%%
\begin{table}
\begin{tabular}{l*{6}{c}r}
                          & $N_p=10^4$	 	& $N_p=10^5$ 	 \\
\hline
$f^H$               	  & $ 0.443 $ & $ 0.140 $ \\
$f^P$ 	  	  	  & $ 0.163 $ & $ 0.090 $ \\
$f^W$ 	  	  	  & $ 0.173 $ & $ 0.086 $ \\
\end{tabular}
\caption{
\label{bump_tail_error}
Comparison of normalized root mean squared errors $e_0$ (\ref{error_e0}) for the raw histogram and for the WBDE and POD methods, for the bump-on-tail instability at $t=149$,
depending on the number of particles.
The simulation with $N_p=10^6$ is used as a reference to compute the error.
}
\end{table}
%%%%%%%%%%%%%%%%%%%%%%%%%%%%%%%%%%%%

\begin{figure*}%[htbp]
%\begin{centering}
{
\includegraphics[width=1.75\columnwidth]{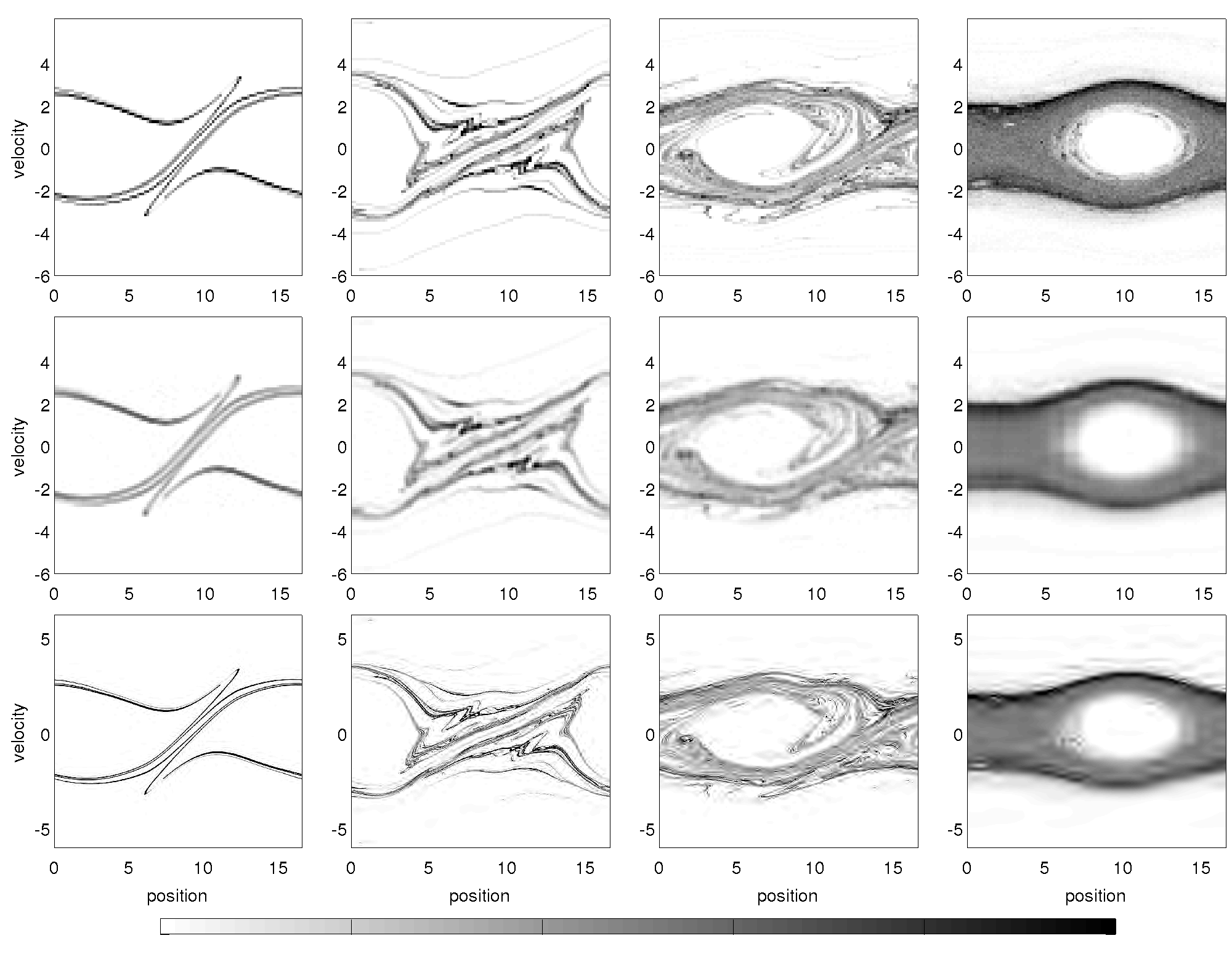}
}
%\end{centering}
\caption
{
\label{two_streams_snapshot}
Contour plots of estimates of $f$ for the two streams instability PIC data at 
times $t=40$, $t=60$, $t=100$ and $t=400$ (left to right).
Histogram method  (first row), WBDE  method (second row), and POD method (third row),
The gray level tone varies uniformly in the interval $[0,A]$, where $A=0.15$, $A=0.08$, $A=0.05$ and $A=0.025$ in the first, second, third and fourth columns respectively.
}
\end{figure*}

The analysis is focused on four stages of the instability corresponding to $t=40$, $60$, $100$, and $400$.
Fig.~\ref{two_streams_snapshot} shows a comparison of the raw histogram, the POD and the WBDE reconstructed particle distribution functions at these four instants.
Grid sizes were $N_g=1024$ for the WBDE estimate, and $N_g=128$ for the two others.
For $t=40$, no noise seems to have affected the particle distribution yet,
therefore a perfect denoising procedure should conserve the full information about the particle positions.
Although WBDE introduces some artifacts in regions of phase space that should contain no particles at all,
it remarkably preserves the global structure of the two streams.
This is possible thanks to the numerous wavelet coefficients close to the sharp features in $f$ that are above the thresholds, in contrast to the bump-on-tail case.
On the next snapshot at $t=60$, the filaments have overlapped and the system is beginning to loose
its memory due to numerical round-off errors.
The fastest filaments still visible on the histogram are not preserved by WBDE, but the most active regions are well reproduced.
At $t=100$, the closeness between the histogram and the WBDE estimate is striking.
To put it somewhat subjectively, one may say that WBDE did not consider most of the rough features present at this stage as 'noise', since they are not removed.
Only with the last snapshot at $t=400$ does the WBDE estimate begin to be smoother than the histogram, suggesting that the nonlinear interaction between particles has introduced randomization in the system.

%%%%%%%%%%%%%%%%%%%%%%%%%%%%
%%%%%%%%%%%%%%%%%%%%%%%%%%%%

The POD method is able to track very well the small and large scale structures of the particle density using a significantly smaller number of modes. In particular, for  $t=40$, $60$, $100$, and $400$ only 
$r=28$, $r=27$, $r=18$, and $r=5$ modes were kept. The decrease of the number of modes with time is a result of the lost of fine scale features in the distribution function. Despite this, a limitation of the POD method is the lack of a thresholding algorithm to determine the optimal number of modes a priori.

%%%%%%%%%%%%%%%%%%%%%%%%%%%%%%%%%%
\section{Summary and Conclusion}
%%%%%%%%%%%%%%%%%%%%%%%%%%%%%%%%%%

Wavelet based density estimation was investigated as a post-processing tool to reduce the noise in the reconstruction of particle distribution functions starting from discrete particle data. This is a problem of direct relevance to particle-based transport calculations in plasma physics and related fields. In particular, particle methods present many advantages over continuum methods, but have the potential drawback of introducing noise 
due to statistical sampling. 

In the context of particle in cell methods this problem is typically approached using finite size particles. However, this approach, which is closely related to the kernel density estimation method in statistics, requires the choice of a smoothing scale, $h$, (e.g., the standard deviation for Gaussian shape functions) whose optimal value is not known a priori. A small $h$ is desirable to fit as many Debye wavelengths as possible, whereas a large $h$ would lead to smoother distributions. This situation results from the compromise between bias and variance in statistical estimation. 
To address this problem we proposed a wavelet based 
density estimation (WBDE) method that does not require an a priori selection of a global smoothing scale and that its able to adapt  locally to the smoothness  of the density based on the given discrete data.  
The WBDE was introduced in statistics \cite{Donoho1996}. In this paper we extended the method to higher dimension and applied it for the first time to particle-based calculations. The resulting method exploits the multiresolution properties of wavelets, has very weak dependence on adjustable parameters, and relies mostly on the raw data to separate the relevant information from the noise.

As a first example, we analyzed a plasma collisional relaxation problem modeled by stochastic differential equations.
%We have shown that the wavelet representation of the deterministically relevant distribution function is sparse.
Thanks to the sparsity of the wavelet expansion of the distribution function, we have been able to extract the information out of the statistical fluctuations by nonlinear thresholding of the wavelet coefficients.
At late times, when the particle distribution approaches a Maxwellian state, we have been
able to quantify the difference between the denoised particle distribution function 
and its analytical counterpart, thus demonstrating the improvement with respect to the raw histogram.
The POD-smoothed and wavelet-smoothed particle distribution functions were shown to be roughly equivalent in this respect.
These results were then extended to a more complex situation simulated with a $\delta f$ code.
Finally, we have turned to the Vlasov-Poisson problem, which includes interactions between particles
via the self-consistent electric field.
The POD and WBDE methods were shown to yield quantitatively close results in terms of mean squared error
for a particle distribution function resulting from nonlinear saturation after occurrence of a bump-on-tail instability.
We have then studied the denoising algorithm during nonlinear evolution after the two-streams instability starting from two counter-streaming cold electron beams.
This initial condition violates the decorrelation hypothesis underlying the WBDE algorithm, and thus offers a good way to test its robustness regarding this aspect.
The WBDE method was shown to yield qualitatively good results without changing the threshold values.
%The choice of the thresholds is based on the test particle superposition principle, which is valid
%when the simulated particles do not collide at all.

%In general, if the plasma is not collisional enough, the denoising scheme that we have proposed may need to be adjusted to explicitly take
%into account correlations between particles \cite{Krommes2007}.
One  limitation of the present work comes from the way denoising quality is measured.
We have considered the quadratic error on the distribution function $f$ as a first indicator of the quality of our denoising methods.
However, it may be more relevant to compute the error on the force fields, which determine the evolution of the simulated plasma.
These forces depend on $f$ through integrals, and statistical analysis of the estimation of $f$ using weak norms,
like was done in \cite{Victory1991} in the deterministic case, could therefore be of great help to obtain
threshold parameters more efficient than those considered in this study.
The computational cost of our method scales linearly with the number of particles and with the grid resolution.
Therefore, WBDE is an excellent candidate to be performed at each time step during the course of a simulation.
Once the wavelet expansion of the denoised particle distribution function is known,
it is possible to continue using the wavelet representation to solve the Poisson equation \cite{Jaffard1992} and to compute the forces. 
The moment conservation properties that we have demonstrated in this paper should mitigate the unavoidable dissipative effects implied by the smoothing stage.
In Ref.~\cite{McMillan2008}, a dissipative term was introduced in a global PIC code to avoid unlimited growth of particle weights in $\delta f$ codes,
and this was shown to improve long time convergence of the simulations.
It would be of interest to assess if the nonlinear dissipation operator corresponding to WBDE has the same effect.
%Demonstrating this using a toy plasma model will be a natural continuation of the present work.

%Pushing the denoising even further, one could periodically resample the particle distributions in a quiet way from the denoised $f^W$, in the spirit of \cite{Denavit1972}.

%The same kind of convergence study could of course also be 

%In hot plasmas, important collective phenomena occur at the Debye length scale which,
%at the present stage of our theoretical understanding, have to be explicitely resolved by the numerical simulation.
%Note that plasma simulations using finite size particles involves two steps: going from the particle data to the grid data, and then computing the forces on the particles based on the fields.
%Our proposed method addresses only the first step.
%In \cite{Besse2007}, a wavelet-based multiresolution method is proposed to solve the Vlasov equation in a semi-Lagrangian way.

\subsection*{Acknowledgements}
We thank D. Spong for providing the DELTA5D Monte-Carlo guiding center simulation data in Fig.6, originally published in Ref.~\cite{delCastillo2008}.
We also  thank Xavier Garbet for his comments on the paper and for pointing out
several key references.
MF and KS acknowledge financial support by ANR under contract M2TFP, M\'ethodes multi\'echelles pour la turbulence dans les fluides et les plasmas.
DCN and GCH acknowledge support  from the Oak Ridge National Laboratory, managed by UT-Battelle, LLC, for the U.S.
Department of Energy under contract DE-AC05-00OR22725. DCN also gratefully acknowledges the support and hospitality of the \'Ecole Centrale de Marseille for the three, one month visiting positions during the elaboration of this work. 
%DCN thanks \'Ecole Centrale Marseille for a 3 month visiting position during the preparation of this paper.
This work, supported by the European Communities under the contract of Association between EURATOM, CEA and the French Research Federation for fusion studies, was carried out within the framework of the European Fusion Development Agreement. The views and opinions expressed herein do not necessarily reflect those of the European Commission.
%RNVY, MF and KS acknowledge financial support by the Euratom-CEA association.

%% Bibliography
%

\end{document}